\documentclass{article}

\usepackage[utf8]{inputenc}
\usepackage{geometry}
\usepackage{times}
\usepackage{setspace}
\usepackage{hyperref}
\usepackage{authblk}
\usepackage{graphicx} 
\usepackage{rotating}
\usepackage{multirow}
\usepackage{natbib}
\usepackage{float}

\newcommand{\logg}{$\log$g}
\newcommand{\kep}{{\it Kepler}}
\newcommand{\gaia}{{\it Gaia}}

\newcommand{\etl}{{\it et al.\,}}

\title{A variable star population in the open cluster NGC\,6791 observed by the \kep\ spacecraft}

\author{\small S.\,Sanjayan$^{1,2}$, A.S.\,Baran$^{1,3,4}$, P.\,N\'emeth$^{1,5,6}$, K.\,Kinemuchi$^{7,8}$, J.\,Ostrowski$^{1}$ and S.K.\,Sahoo$^{1,2}$\\
$^{1}$ARDASTELLA Research Group, Institute of Physics, Pedagogical University of Cracow, \\ ul. Podchor\c{a}\.zych 2, 30-084 Krak\'ow, Poland\\
$^{2}$Centrum Astronomiczne im. Miko{\l}aja Kopernika, Polskiej Akademii Nauk, ul. Bartycka 18, 00-716 Warszawa, Polska\\
$^{3}$Embry-Riddle Aeronautical University, Department of Physical Sciences, Daytona Beach, FL\,32114, USA\\
$^{4}$Department of Physics, Astronomy, and Materials Science, Missouri State University, Springfield, MO\,65897, USA\\
$^{5}$Astroserver.org, F\H{o} t\'er 1, 8533 Malomsok, Hungary\\
$^{6}$Astronomical Institute of the Czech Academy of Sciences, Fri\v{c}ova 298, CZ-251\,65 Ond\v{r}ejov, Czech Republic\\
$^{7}$Department of Astronomy, New Mexico State University, Box 30001, MSC 4500, Las Cruces, NM 88003, USA\\
$^{8}$Apache Point Observatory, 2001 Apache Point Road, P.O. Box 59, Sunspot, NM 88349-0059
}
\date{April 15, 2022}

\begin{document}

\maketitle

\begin{abstract}
We present the list of variable stars we found in the \kep\ superstamp data covering approximately 9\,arcminutes from the central region of NGC\,6791. We classified the variable stars based on the variability type and we established their cluster membership based on the available \gaia\ Early Data Release\,3 astrometry, by means of the Bayesian Gaussian mixture models. In total we found 278 variable objects, among which 17 binaries, 45 pulsators, 62 rotational and five unclassified variables are cluster members. The remaining 28 binaries, 25 pulsators, 83 rotational, four unclassified and nine unidentified variables are either not members or their membership is not established. In the case of eclipsing binaries we calculated the mid-times of eclipses and derived ephemerides. We searched for eclipse timing variation by means of the observed minus calculated diagrams. Only three objects show significant orbital period variation. Independently of a report published just recently by Colman \etl(2022) we found 119 new variables. We used isochrones calculated within the MIST project and derived the age (8.91\,Gyr), average distance (4134\,pc) and iron content [Fe/H] (0.26-0.28), of NGC\,6791. Using the cluster members with membership probabilities greater than 0.9, we calculated the distance to the cluster of 4123(31)\,pc, which agrees with the result from our isochrone fitting.\\
{\bf Open clusters and associations : individual: NGC\,6791 -- binaries: general -- Stars oscillations -- Stars : rotation}
\end{abstract}

\section{Introduction}
NGC\,6791 has been first described as a metal rich cluster by Baade\,(1931) and listed as an old open cluster by King\,(1964). The authors provided no age estimations. Kinman\,(1965) presented a detailed comparative study of color\,--\,magnitude diagrams\,(CMD) of NGC\,6791 along with other two open clusters, M\,67\,(4\,Gyr) and NGC\,188\,(6.8\,Gyr). From the first photometric observations in the B-V color, Harris and Canterna\,(1981) determined a reddening of E(B-V)\,=\,0.13\,mag. According to the recent studies NGC\,6791 is 7\,--\,9\,Gyr old (Chaboyer \etl1999, Carraro \etl2006, Basu \etl2011), and it has a mass of around 4\,000\,M$_{\odot}$\,(Kaluzny and Udalski\,1992, Carraro \etl2006, Platais \etl2011, Tofflemire \etl2014). The cluster is located $\sim$8000\,pc from the Galactic center and 1000\,pc above the Galactic plane. According to some hypotheses the cluster may have formed in the bulge of the Galaxy and radially migrated to its current location (Jilkova \etl2012, Villanova \etl2018). The distance to the cluster is approximately 3\,614\,pc, which was estimated for the first time by Stetson \etl(2003) from the de-reddened distance modulus of (m-M)$_0\approx$12.79\,mag. The authors derived E(B-V)\,=\,0.09\,mag. According to Villanova \etl(2018) NGC\,6791 is a super metal-rich cluster with [Fe/H]\,=\,+0.3\,--\,+0.4. Geisler \etl(2012) showed that NGC\,6791 has multiple stellar populations, which makes the cluster chemically peculiar. NGC\,6791 has an anomalous horizontal branch with a red clump\,(RC) region. Liebert \etl(1994) found a group of extreme horizontal branch members using spectrophotometry of blue targets observed by Kaluzny and Udalski\,(1992). The age of the cluster predicts that it should have a rich population of cooling white dwarfs, hence Bedin \etl(2005) observed the cluster using the Hubble Space Telescope up to m$_\mathrm{F606W}\approx$\,28.5\,mag. They found the white dwarf luminosity function to give a peak at 27.4\,mag, which corresponds to an age of 2.5\,Gyr. Such an estimate does not agree with the age derived from the main sequence\,(MS) or red giant branch\,(RGB) population (Chaboyer \etl1999, Carraro \etl2006). Thus far, these studies show the cluster is very unusual.
A more detailed study is required to constrain the age and metal abundances for understanding the formation and evolution of NGC\,6791. A clear picture of the cluster could be achieved by deriving an entire population of variable stars and analysis of components of the stars to find their ages and chemical abundances.

NGC\,6791 has been a subject of extensive search for variable stars. Kaluzny and Udalski\,(1992) and Kaluzny and Rucinski\,(1993) did an extensive photometric survey finding 17 variable stars which includes 8 contact binaries, two blue stragglers and one binary consisting of a hot subdwarf B star. Rucinski \etl(1996) found three detached binaries and one cataclysmic variable (CV) star exhibiting a three day outburst. As a part of search for planets in stellar clusters, Mochejska \etl(2002) found 47 new low amplitude variable stars. The authors reported several BY\,Dra type and two outbursting CV stars, confirming the CV found by Rucinski \etl(1996). Mochejska \etl(2003) reported seven new variable stars with a long and periodic flux variation. Kaluzny\,(2003) found four new variable stars by reanalyzing archived data from Kaluzny and Rucinski\,(1993). A search for transiting events by giant planets reported by Bruntt \etl(2003) yielded 22 new low amplitude objects along with 20 previously known variable stars. Using a high precision time-series photometry, Hartman \etl(2005) detected 10 new variable stars including one $\delta$-Scuti type star and 8 contact binaries. Mochejska \etl (2005) detected 14 more variable stars and reported 9 eclipsing binaries. Using a high precision photometry in the Johnson V band, de Marchi \etl(2007) detected 260 variables in the cluster area, although not all stars are members of the cluster.

From the launch in 2009, for almost 10 years the \kep\ spacecraft has served mankind by providing very precise and almost continuous photometric measurements (Koch \etl 2010). The \kep\ has observed more than five hundred thousand stars during its entire mission time. The \kep\ mission was completed in two phases. During the first mission, \kep\ observed 0.25\% of the sky in the direction of Cygnus and Lyra constellation for 1460\,days. The mission was reborn as K2\,(second mission) after the second reaction wheel failed. K2 mission made 80\,day observing campaigns along the ecliptic equator, which lasted 1695\,days (Howell \etl 2014). During both missions, the observations were obtained using two different exposures, 30 minutes for the long cadence\,(LC) and 1 minute for the short cadence\,(SC) mode (Koch \etl 2010, Borucki \etl 2010, Caldwell \etl 2010, Thompson \etl 2016). During the first mission, four open clusters were inside the \kep\ field of view, NGC\,6791, NGC\,6819, NGC\,6811 and NGC\,6866. Two of the open clusters, NGC\,6791 and NGC\,6819, were observed by using the so-called LC superstamps.

Recently, Colman \etl(2022) presented light curves of KIC stars obtained from the \kep\ superstamp data. The authors used an image subtraction method to derive light curves of all \kep\ cataloged targets. They identified variability in 239 out of 5342 stars they extracted light curves of. The number of new variables is not given. We stress that our work has been performed simultaneously to, yet independently from, Colman \etl(2022) and contains additional analysis. By comparing our results with results of Colman \etl(2022), we have noticed that the authors applied a very strong detrending policy removing either eclipses or out-of-eclipse variations in binaries or variations in other objects that we claim to be variables.

In Section 2 of this paper, we present a brief description of the \kep\ data and method used for obtaining the light curves of variable objects. In Section 3, we present a spectroscopic study of the variable stars found in this project using either archived spectra from public surveys or our own data. In Section 4, we describe the method of deriving the membership probabilities of our new variable star findings. In Section 5, we report individual variable star cluster members divided into variability classes. The field variable counterparts are listed in the Tables 5--8. In Section 6, we present the result of isochrone fitting.

\section{Kepler Photometry}
We downloaded the \kep\ superstamp data of NGC\,6791 from the Mikulski Archive for Space Telescopes\,(MAST\footnote{{https://archive.stsci.edu/}}). The data are 20\,x\,100 pixel boxes piled up in two contiguous 10 box stacks. The field of view of all pixels is 800\,x\,800\,arcseconds and covers the most central part of the cluster. The superstamps data are collected in the LC mode. The pixel scale of an individual square pixel is 4\,arcsec. The data have been collected over 1460\,days and are split into 18 quarters.

We searched for a flux variation by extracting fluxes for all time stamps in individual pixels for each of the quarters Q\,2\,--\,5. Then, a Fourier transform of the time-series data was performed in each pixel and each quarter separately. The pixels showing peaks (representing signal) in the amplitude spectra were selected. Signals that were identified with artifacts, either reported by Baran\,(2013) or those found in this project, were discarded. We combined all contiguous pixels showing the same signal and defined an optimal aperture of pixels. To keep the solar cells exposed to the sunlight, every quarter the spacecraft rolled 90\,degrees, hence, with each quarterly rotation of the spacecraft, our targets landed on different CCD chips. This positioning caused different target images, and consequently, different optimal apertures (Bryson \etl 2010). Fortunately, every four quarters the images and apertures were the same, so we have defined the apertures only in four quarters, i.e. Q\,2\,--\,5, and propagate them to the corresponding quarters (e.g. Q\,2,\,6,\,10,\,14). Next, using the optimal apertures for all targets showing flux variation we used PyKE software (Kinemuchi \etl2012) to pull out the fluxes and correct them for instrumental artifacts by means of Co-trending Basis Vectors. Finally, using our custom scripts, we clipped the data at 4.5\,sigma, detrended using spline fits, and normalized them to {\it parts per thousand} (ppt). The variable stars discovered in our work will be presented in Section\,5.

\section{Spectroscopy}
We searched for spectra in the literature of all variables we detected. We found optical or infrared spectra for 111 objects in the archives of APOGEE (Ahn \etl 2014, Majewski \etl 2017), SDSS (Blanton \etl 2017), LAMOST (Zhao \etl 2012), ESO (Gilmore \etl 2012, Randich \etl 2013), and the HECTOSPEC (Fabricant \etl 2005) surveys. All spectra with T$_{\rm eff}<15\,000$\,K were modeled with interpolated local thermal equilibrium (LTE) synthetic spectra drawn from the BOSZ (Bohlin \etl2017) spectral library to determine the fundamental atmospheric parameters. The BOSZ library was calculated for scaled solar metallicity with carbon and $\alpha$-element enhancement; therefore, individual abundance patterns cannot be investigated with our method. 

Our fitting procedure ({\sc XTgrid}; N\'{e}meth \etl2012) is based on a steepest-gradient chi-square minimizing method, which was originally developed to model hot stars. To improve its performance for cool stars, we added a grid-search preconditioning to the procedure. We step through a set of models to search for the best starting model for the steepest-descent part. Next, the descent part takes over in driving the fit and converges on the best solution. Once a convergence is achieved, the procedure explores the parameter errors by stepping through a set of points around the best solution. If a better solution is found during error calculations, then the procedure returns to the descent part, and hence pushing the solution towards the global minimum. {\sc XTgrid} fits the radial velocity and projected rotation velocity of each spectra along with the stellar surface parameters, such as the effective temperature (T$_{\rm eff}$), surface gravity (\logg) and abundances.

In addition, the procedure accesses photometric data from the VizieR Photometry Viewer\footnote{\url{http://vizier.u-strasbg.fr/vizier/sed/}}, distance data from the Gaia EDR3 database, and extinction values from the NED online services. The spectroscopic surface parameters combined with these measurements allow us to reduce systematics and derive absolute stellar parameters, such as mass, radius, and luminosity. An anti-correlation is observed between T$_{\rm eff}$ and [Fe/H]. Fortunately, the spectral energy distribution (SED) helps in resolving this bias by restricting the T$_{\rm eff}$. Another bias is observed in surface gravity, in particular below T$_{\rm eff}$\,=\,4\,000\,K. At such low temperatures, the spectrum is insensitive to the surface gravity. When the spectral coverage is very limited, we could not determine an accurate value for \logg. We do not report atmospheric parameters for such stars. 

The archival spectroscopic data are very inhomogeneous. Consequently, high resolution spectra (e.g. obtained with ESO instruments) with a short wavelength coverage are more suitable for radial velocity measurements, while low resolution spectra (e.g. from the SDSS and LAMOST surveys) can provide more consistent atmospheric parameters, but less precise velocities. Some ESO spectra cover only 5\,300-5\,600\,\AA\ range at a resolution of R=20\,000, and only weak spectral features are visible. For such spectra, at a relatively low signal-to-noise ratio (SNR), the fitting procedure increases the projected rotation above 100\,km\,s$^{-1}$, which decreases the radial velocity accuracy. In general, low SNR spectra limit our analysis the most, while crowding in dense stellar fields and a limited spectral coverage affects the parameter determination.

\section{Cluster membership}
We used \gaia\ astrometry to determine the membership probabilities of all variable stars we found. We used five parameters, i.e. equatorial coordinates $\alpha$ and $\delta$, proper motions $\mu_{\alpha}$ and $\mu_{\delta}$, and parallax $\pi$ (further called five astrometric parameters). First, we adopted/estimated mean values of these parameters. The cluster center has been taken from Kamann \etl(2019) to be at $\alpha_{\rm 2000}$\,=\,19:20:51.3 and $\delta_{\rm 2000}$\,=\,+37:46:26. Next, we downloaded \gaia\ Early Data Release\,3 (EDR3) (\gaia\ collaboration \etl2016, 2021) data for all stars within a tidal radius of 23\,arcmin (Platais \etl2011). The area contains 36\,647 targets accessible to our analysis; however, we filtered out dubious targets with parallaxes to be negative or greater than 1\,arcsec or relative uncertainties for any of the proper motion or parallax values to be greater than 50\%. A cluster environment, particularly toward the center, is very dense, which can lead to unrealistic or imprecise estimates of these three parameters ($\mu_{\alpha}$, $\mu_{\delta}$, $\pi$). In addition, we limited our sample to targets for which the zero point offset corrections of parallax have been applied (Lindegren \etl2021). After filtering and correcting for the parallax zero offset, we ended up with 11\,466 targets.

To determine the membership probabilities, we used the Bayesian Gaussian Mixture Models (GMM) using {\it scikit--learn} python toolkit (Pedregosa \etl2011). The GMM assumes each data point to be a combination of finite Gaussian functions, in which the number of these functions is determined using a variational Bayesian inference model with Dirichlet process prior (Ferguson 1973). We performed 10\,000 iterations using the Expectation\,--\,Maximization algorithm (Dempster \etl1977), and we derived membership probabilities for each target in our sample based on all five astrometric parameters. In the case of targets that we found variables in the superstamp area, we estimated their membership probabilities regardless of precision of their five astrometric parameters. If the uncertainties were larger than 50\%, we considered corresponding parameters to be error-free, while the negative parallaxes were ignored and only four astrometric parameters were used.

To strengthen the probability, the radial velocity of individual stars can also be used, however they need to be corrected for the effects of binarity, rotation, and pulsations. Different instruments differ in instrumental calibration which often bias the radial velocity (RV) estimates. Since we did not conduct one single survey that could provide us with consistent RV estimations, we decided not to use RVs for the membership analysis. Since binarity and rotation affects the measurement of intrinsic motion, we expect the RVs will be random values and, as it will be seen in Section\,5, the values in Tables\,1--3, confirm our suspicion. We expected the most consistent estimates for single solar-like pulsators, since their oscillation motion on the surface is of a very small amplitude. In fact, in only three cases the RVs are far from the average cluster value (-47.46$\pm$1.08 km/s, Carrera \etl2019), since the stars may belong to binary systems. On the other hand, RVs that are consistent indicate that the stars are likely single or the orbital motion (if any) is very slow or the spectra have been taken when both stars were aligned with the observer's line of sight. The RVs of the solar-like stars that are unlikely to be members (Table\,6) are not close to the cluster average and seem to confirm their field membership, unless they are in binaries.

\section{A zoo of variable stars}
In total, we found 278 variable objects in the superstamp area. Our sample contains cluster members as well as foreground and background stars. In Section\,4, we provided details on a membership analysis. Our prime focus is on the members of NGC\,6791. The non-members and objects with unknown membership status, as a consequence of a lack of the \gaia\ astrometry, are listed in the Table 5-8. Their variability is classified the same way as for the cluster members.

\begin{figure}[H]
    \centering
    \includegraphics[width=0.75\textwidth]{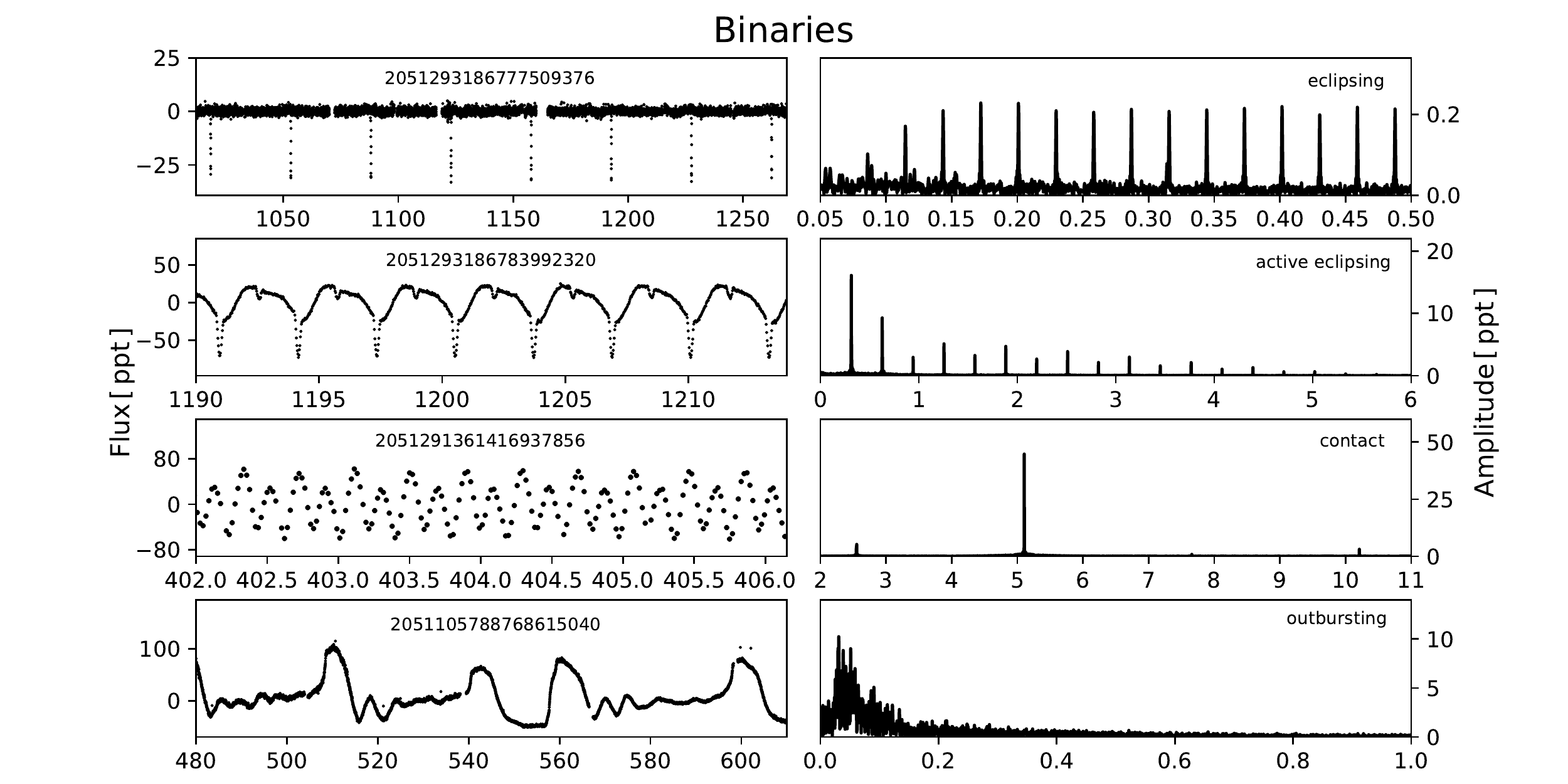}
    \includegraphics[width=0.75\textwidth]{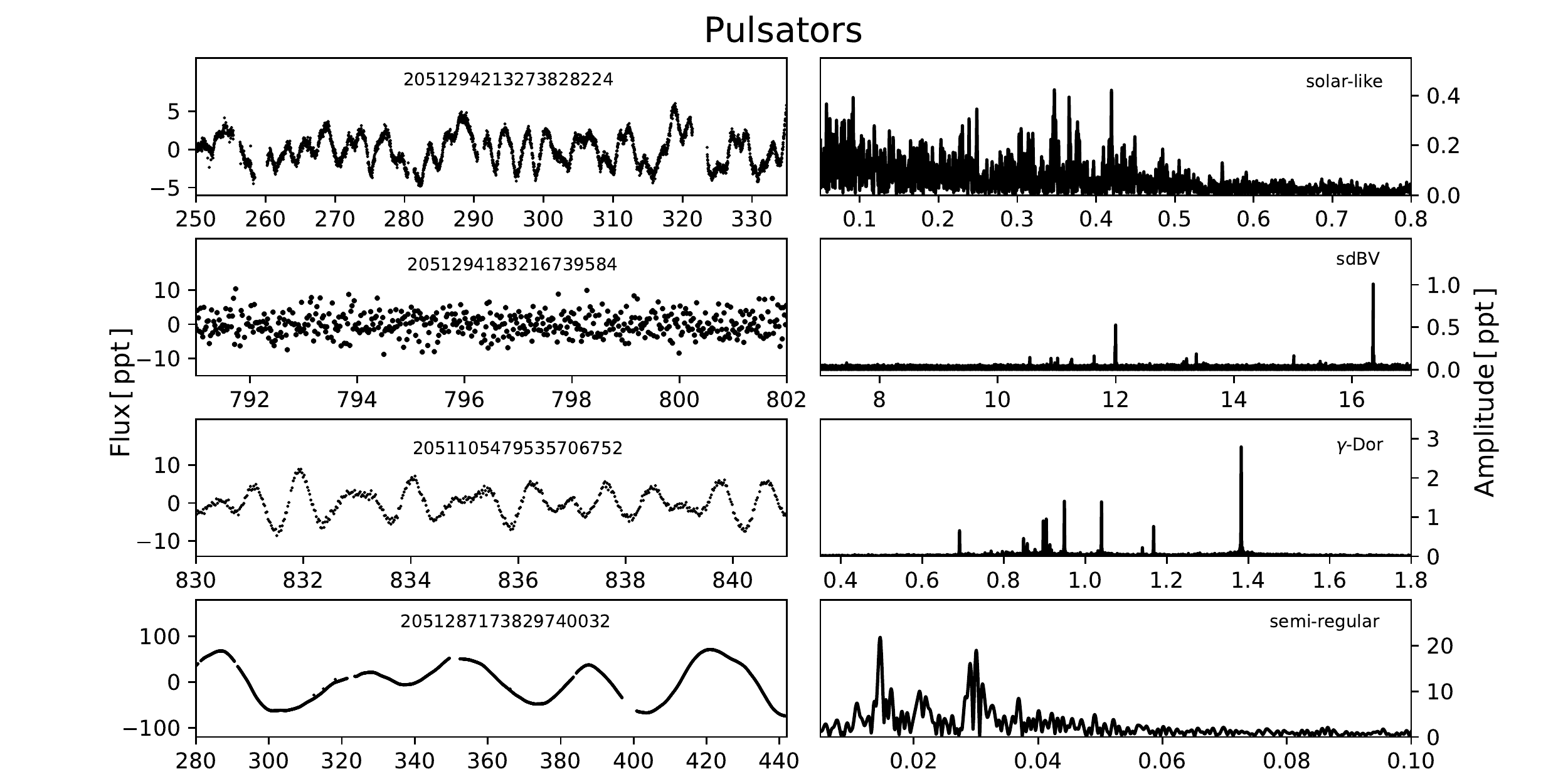}
    \includegraphics[width=0.75\textwidth]{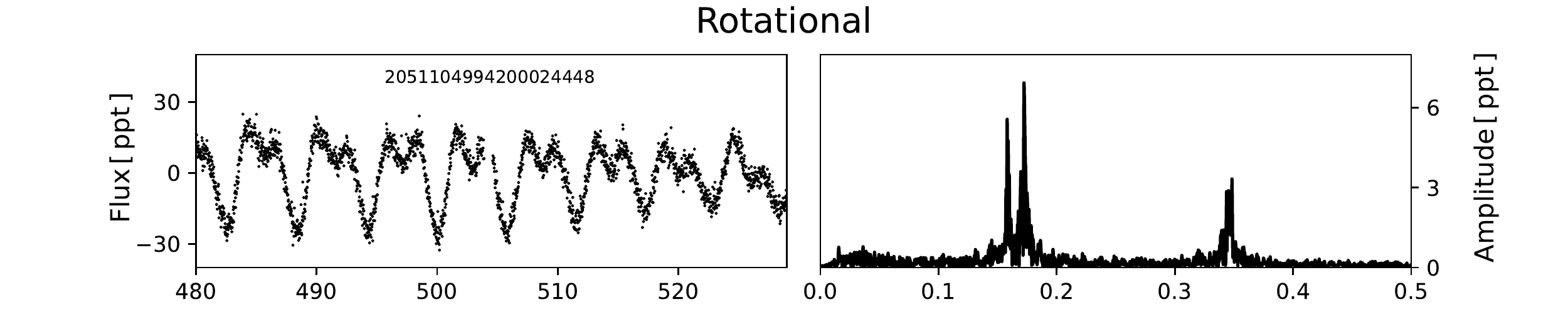}
    \includegraphics[width=0.75\textwidth]{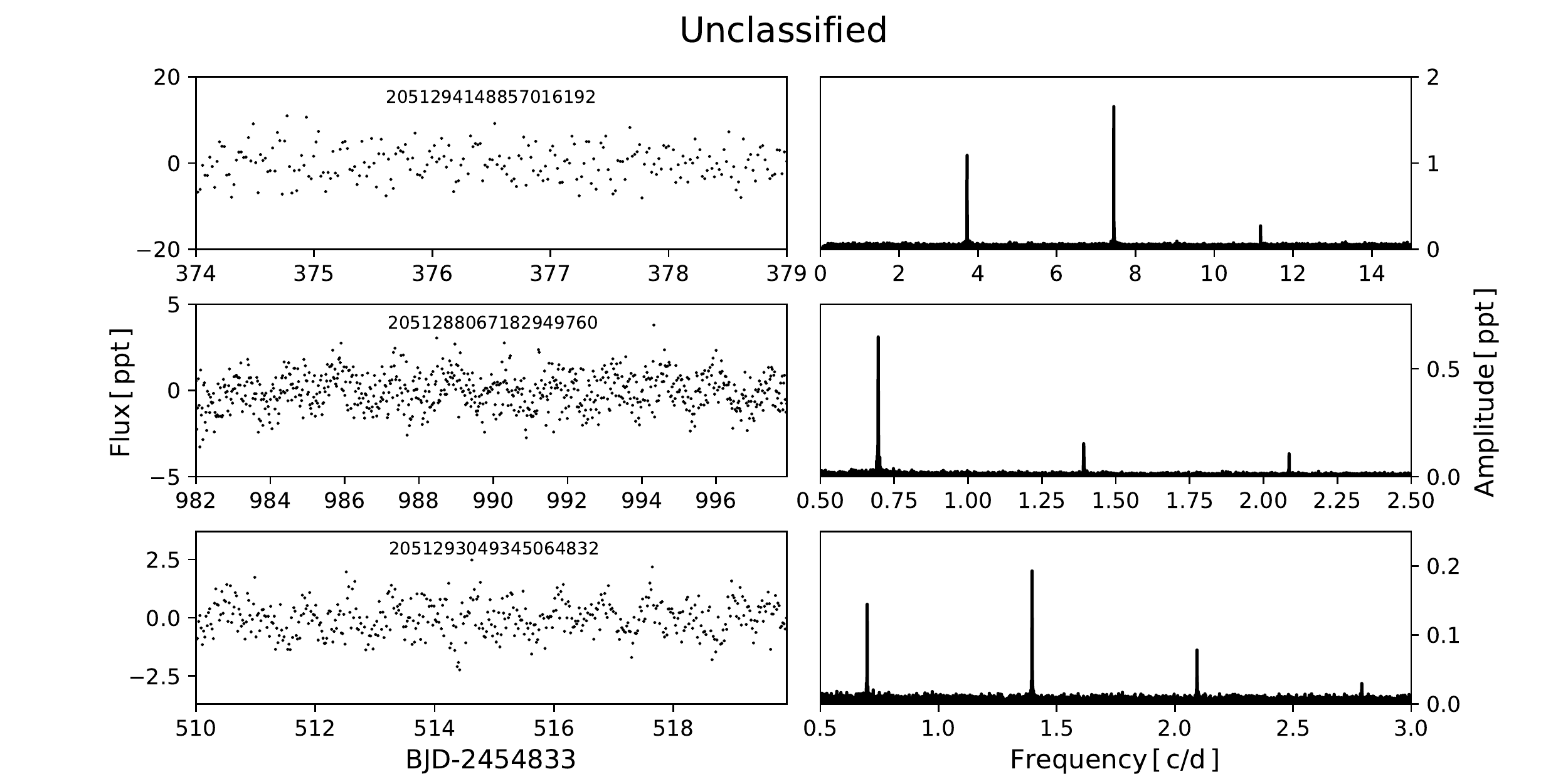}
    
    \caption{Examples of light curves and corresponding amplitude spectra of a zoo of variable stars in the open cluster NGC\,6791.}
    \label{fig:1}
\end{figure}

Based on flux variations, we classified the stars into three main variability types, i.e. eclipsing, pulsating, and rotating stars. The first two types are further split into specific classes. Five stars remained unclassified. Their light curves show variations that we are unable to unambiguously identify as one of the three types listed. These objects show flux variations which can origin in {\it e.g.} a reflection effect, ellipsoidal variation or a rotation of a spotted star. These stars have typically low amplitude flux variations. In Fig.\,1 we present examples of light curves and their corresponding amplitude spectra for each type and a selection of classes of variable stars we found.

\subsection{Binary systems}
We selected binary stars with sharp eclipses typical for semi- and detached systems. Some eclipsing systems show additional out-of-eclipse variation, which can be caused by a chromospheric activity, and we call them "active" eclipsing. We identified contact systems, which are characterized by a continuous flux change and typical for W\,UMa stars. Another class contains outbursting stars, which we associate with binaries experiencing a rapid mass transfer causing sudden eruptions, e.g. novae, dwarf novae, nova-like variables. We stress that our classification is not based on radial velocities. Some of the stars may not be classified correctly, e.g. a smoothly continuous and small amplitude flux changes may be misidentified with rotational variables. However, the flux change over the course of observations is not modulated (see explanation in Section\,5.3) or they can be long-period pulsating stars. In Fig.\,2 we present the phased light curves of three stars that we consider new discoveries. The sample includes all the classes of binary stars we identified in the superstamp data. We found 17 binaries to be cluster members (Table\,1), 28 binaries are field objects, including two binaries for which we could not establish membership due to the lack of \gaia\ astrometry data (Table\,5). For binary systems their membership has been derived based on all five astrometric parameters. Majority of binaries in the cluster are main sequence (MS) stars with just two exceptions, assuming the position of the latter in the CMD is correct. \gaia\,EDR3 2051105720053889536 is a post-MS star on its early ascent of the red giant branch (RGB), while \gaia\,EDR3 2051293186783992320 is located below the RGB, which can be explained by an incorrect color index or pre-MS evolutionary status. Among the member counterparts, five are eclipsing, six active eclipsing, five contact, and one outbursting stars.

\begin{sidewaystable}
\caption{List of binary stars that are cluster members.}
\label{tab:cluster_binaries}
\centering
\resizebox{\columnwidth}{!}{
\begin{tabular}{clllccclllll}
\hline\hline
\multicolumn{1}{c}{\multirow{2}{*}{\gaia\,EDR3}} & \multicolumn{1}{c}{\multirow{2}{*}{KIC}} & \multicolumn{1}{c}{P$_{\rm orb}$} & \multicolumn{1}{c}{T$_0$} & \multicolumn{1}{c}{G} & \multirow{2}{*}{CMD} &  \multirow{2}{*}{HRD} & \multicolumn{1}{c}{T$_{\rm eff}$} & \multicolumn{1}{c}{\multirow{2}{*}{\logg}} & \multicolumn{1}{c}{RV} & \multicolumn{1}{c}{\multirow{2}{*}{[Fe/H]}} & \multicolumn{1}{c}{\multirow{2}{*}{Ref}}\\
&& \multicolumn{1}{c}{[days]} & \multicolumn{1}{c}{[BJD]} & \multicolumn{1}{c}{[mag]} &&& \multicolumn{1}{c}{[K]} && \multicolumn{1}{c}{[km/s]} &&\\
\hline\hline
\multicolumn{12}{c}{eclipsing}\\
2051287203888386688 & 2436378\,$^{10L}$ & 8.531633(8) & 2\,454\,969.2757(8) & 19.031 & MS$^+$ & -- & -- & -- & --& -- & -- \\
{\bf 2051288372129693440} & 2569175\,$^{X}$ & 43.499395(21) & 2\,455\,041.61331(38) & 18.383 & MS & RGB & 5\,620(40)	& 3.517(18) & -65.8(4) & -0.67(9) & 2$^*$ \\
2051292980625561216 & 2437149\,$^{2S,6L}$ & 18.7986285(37) & 2\,454\,971.02807(16) & 17.480 & MS & MS+MS & 5\,100(900) & 4.5(4) & -80.9(9)  & -0.25(25)  & 2$^*$\\
{\bf 2051293083698347008} & 2437482\,$^{X}$ & 17.551175(14) & 2\,454\,966.9452(7) & 18.210 & MS & RGB & 4\,860(50) & 3.042(20) & -67.5(5) & -0.48(15) & 1$^*$ \\
{\bf 2051293186777509376} & 2437041\,$^{X}$ & 34.859569(14) & 2\,454\,979.90568(33) & 18.055 & MS & MS & 6\,360(40) & 4.58(7) & -57.0(8) & -0.42(11) & 2$^*$ \\
\hline
\multicolumn{12}{c}{active eclipsing}\\
2051105342091761536 & 2437783\,$^{4L}$ & 7.453112(10) & 2\,454\,964.7652(11) & 18.391 & MS & BS & 7\,160(50) & 3.60(8)	& -20.1(6) & -0.82(9) & 2$^*$ \\
2051105548255223680 & 2438490\,$^{4L}$ & 3.3157657(10) & 2\,454\,967.85556(27) & 17.887 & MS$^+$ & HB & 5\,410(90)& 2.87(7) & -62.2(12) & -0.68(15) & 2$^*$ \\
2051105784476572032 & 2438061\,$^{1S,12L}$ & 4.8858826(16) & 2\,454\,965.01144(28) & 17.721 & MS & HB & 5\,220(30) & 2.197(10) & -46.5(45) & 0.42(22) & 1$^*$ \\
2051293186783992320 & 2437060\,$^{6L}$ & 3.1871038(8) & 2\,454\,969.04204(21) & 16.989 & RGB & MS & 5\,270(20) & 4.253(45) & -59.6(7) & -0.445(47) & 1$^*$ \\
2051294423734919552 & 2569494\,$^{1S,5L}$ & 1.5232747(12) & 2\,454\,965.3339(7) & 17.288 & MS & MS & 5\,580(50) & 4.06(6) & -116(8) & -0.36(11) & 1$^*$ \\
2051295351447975168 & 2570480\,$^{X}$ & 0.7349394(16) & 2\,454\,964.9628(18) & 19.340 & MS$^+$ & HB & 5\,870(160) & 2.34(11) & 41.2(30) & -0.14(7) & 2$^*$ \\
\hline
\multicolumn{12}{c}{contact}\\
2051105543955885696 & 2438413\,$^{X}$ & 0.31754621(10) & 2\,\,454\,964.63952(27) & 18.281 & MS & RGB & 5\,380(100) & 3.347(38) & -34.70(13) & -0.631(46) & 2$^*$ \\
2051105720053889536 & 2438148\,$^{X}$ & 0.29468323(11) & 2\,454\,964.79285(30) & 16.612 & RGB & -- & -- & -- & -- & -- & --\\
2051291361416937856 & 2568950\,$^{X}$ & 0.3917614(15) & 2\,454\,964.91341(33) & 17.555 & MS & MS/RGB & 5\,300(20)& 3.926(20) & -39.50(10) & -1.131(26) & 1$^*$ \\
2051293324222974976 & 2569965\,$^{X}$ & 0.325587259(49) & 2\,454\,964.71162(13) & 17.510 & MS$^+$ & MS & 5\,380(20)& 4.46(8) & -63(6) & -0.36(8) & 1$^*$ \\
2051294114497255936 & 2569630\,$^{X}$ & 0.31265938(26) & 2\,454\,964.7815(7) & 16.964 & MS & MS & 5\,620(50)& 4,030(56) & -41(2) & -0.14(9) & 1$^*$ \\
\hline
\multicolumn{12}{c}{outbursting}\\
2051105788768615040  & 2438114\,$^{4L}$ & 32.27 & --& 18.192 & blue of MS$^+$ & blue of MS  & 27\,000(5\,000)	& 7.3(6)	& -34(25) & -1(1) & 1$^*$ \\
\hline\hline
\end{tabular}
}
\flushleft
\footnotesize{The stars that we consider newly detected variables are marked in {\bf bold}. The superscripts in the KIC column denote if the data are available on the MAST: X stands for no data, L and S stand for the LC and SC data, respectively, and the numbers represent a total number of quarters the date were taken in. CMD refers to a position in the color-magnitude diagram (MS - main sequence, RGB - red giant branch, BS - blue straggler, AGB - asymptotic giant branch, RC- red clump, HB - horizontal branch, EHB - extreme horizontal branch). HRD refers to a position in the T$_{\rm eff}$, $\log g$ diagram. The $^+$ symbol indicates that {\it not} all five astrometric parameters were used in the membership analysis. The Ref column refers to the source of the spectra, i.e. 1 - HECTOSPEC, 2 - ESO, 3 - LAMOST, 4 - SDSS and 5 - APOGEE. The $^*$ indicates that the atmospheric parameters are derived in this work.
HECTOSPEC - https://oirsa.cfa.harvard.edu, ESO - http://archive.eso.org, APOGEE - https://www.sdss.org/dr16/irspec, LAMOST - http://dr6.lamost.org, SDSS - http://skyserver.sdss.org/dr16}
\end{sidewaystable}

\begin{figure}[H]
\centering
\includegraphics[width=0.7\textwidth]{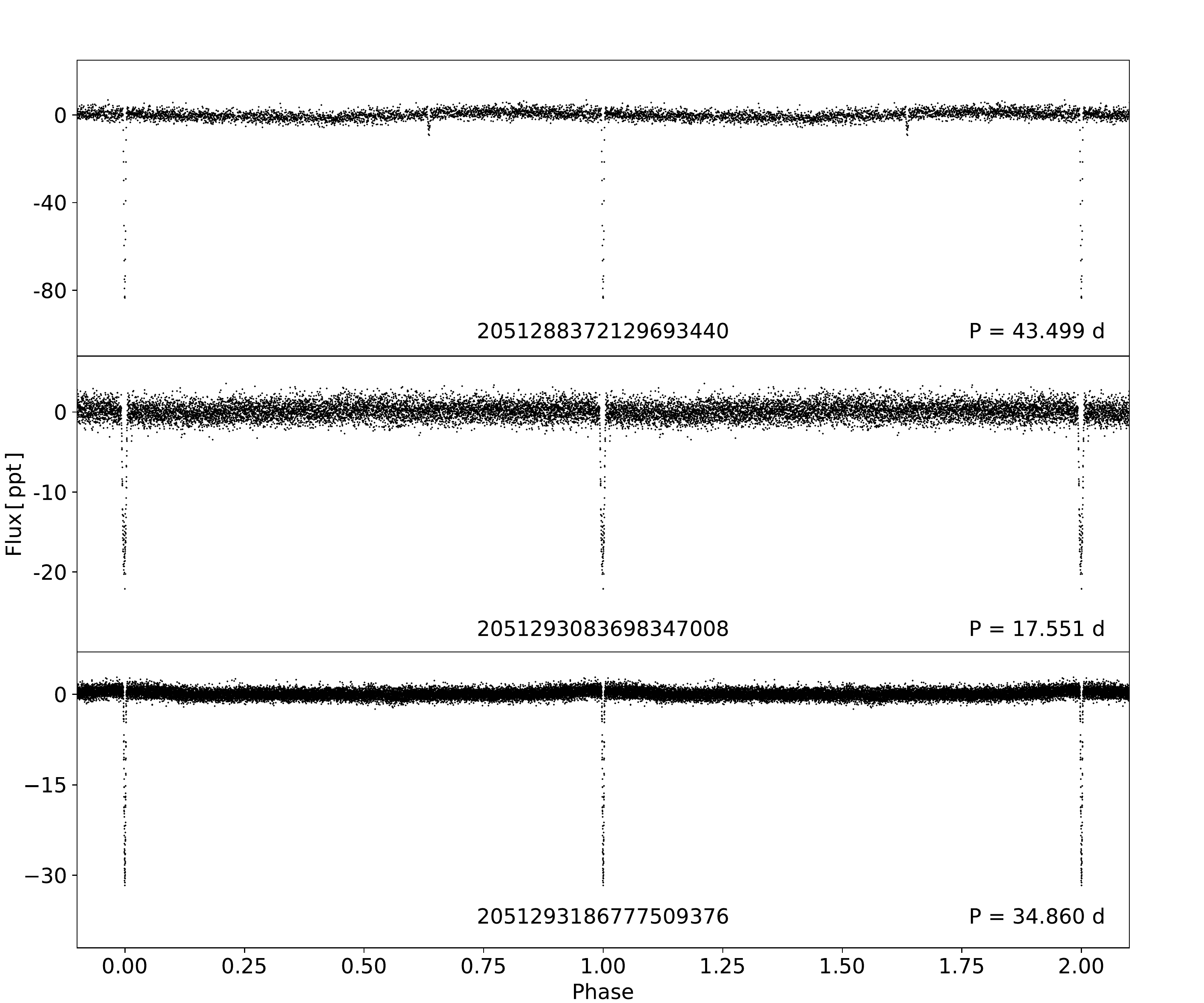}
\caption{Phased light curves of three newly discovered binary star members of NGC\,6791.}
\label{fig:2}
\end{figure}

Since the objects considered in this section are binaries, magnitudes and color indices used in the top panel of Fig.\,3 and T$_{\rm eff}$ and \logg\ in the top panel of Fig.\,6 (this figure will be explained in Section\,6) may be averages of all components in a specific binary system and may not indicate a proper location of individual stars in the CMD. If the average location in the CMD diagram is in the well defined regions, e.g. MS or RGB, we can expect the components are of similar properties (or a secondary component is not detectable in our data), and the average is not far from single star values (unless the binary components are much different but the shift is only along a given evolutionary stage, e.g. along the RGB branch). If the location is outside those regions, e.g. below MS or RGB, then the individual components in a binary system are not alike, e.g. WD + MS, and the average will be somewhere between a WD track and the MS (as is the case for \gaia\,EDR3 2051105788768615040). For this same reason, the RV of binaries is typically different from the average RV of a cluster. We expect the orbital RV to be dominating. There are only one exception with the RV being close to the average cluster value. In the case of \gaia\,EDR3 2051105784476572032 the spectrum could have been taken during the eclipse or it is a spectroscopic single-line binary.

\begin{figure}[H]
\centering
\includegraphics[width=0.7\textwidth]{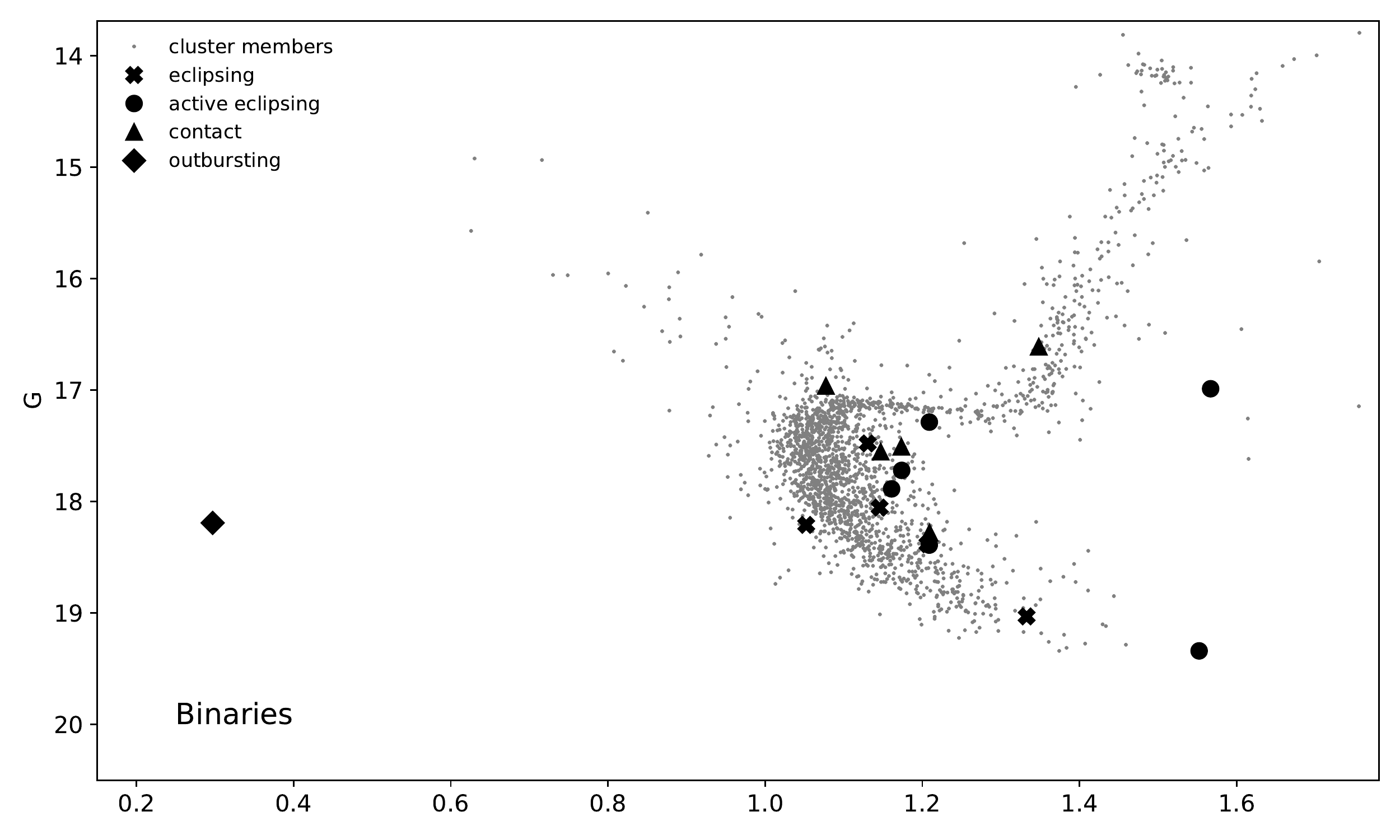}
\includegraphics[width=0.7\textwidth]{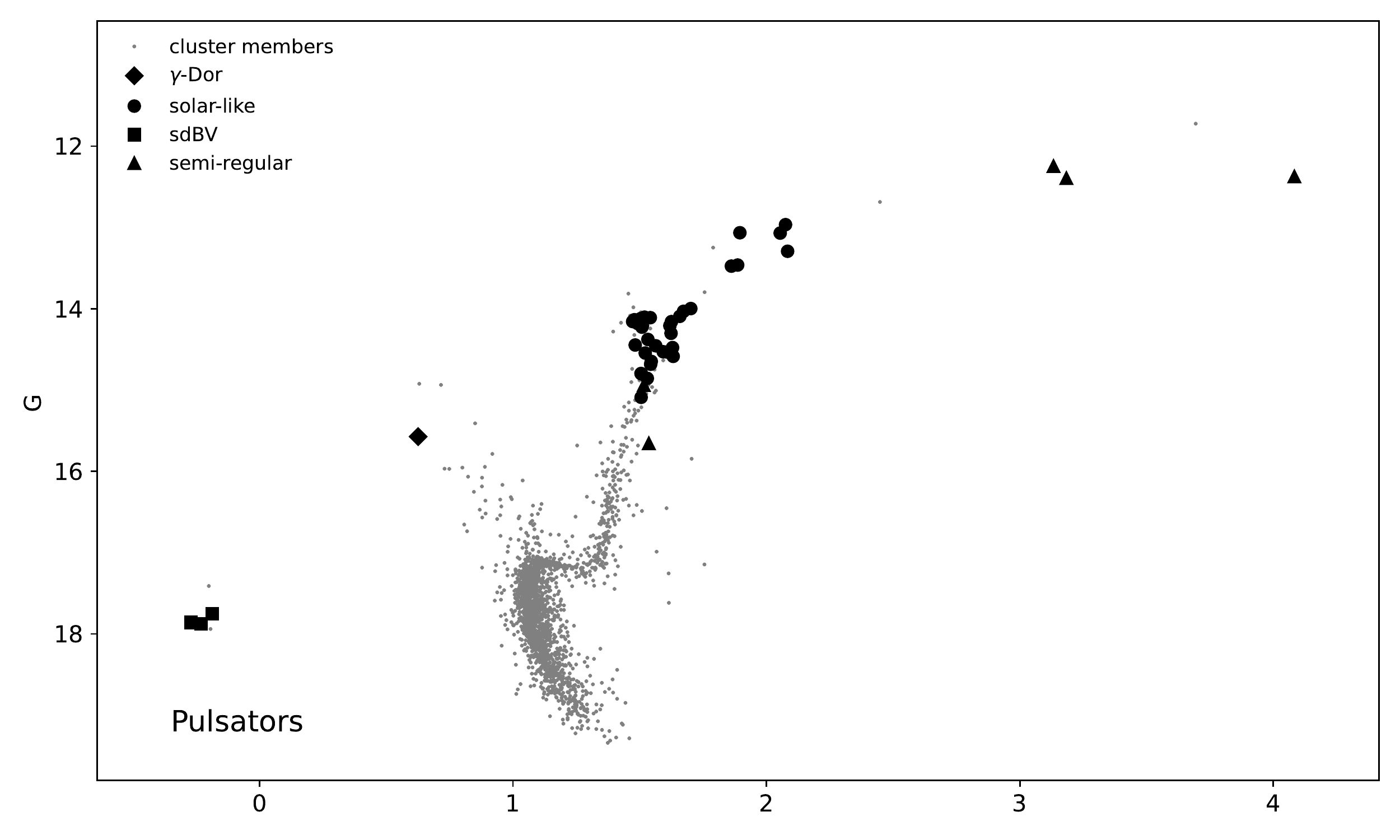}
\includegraphics[width=0.7\textwidth]{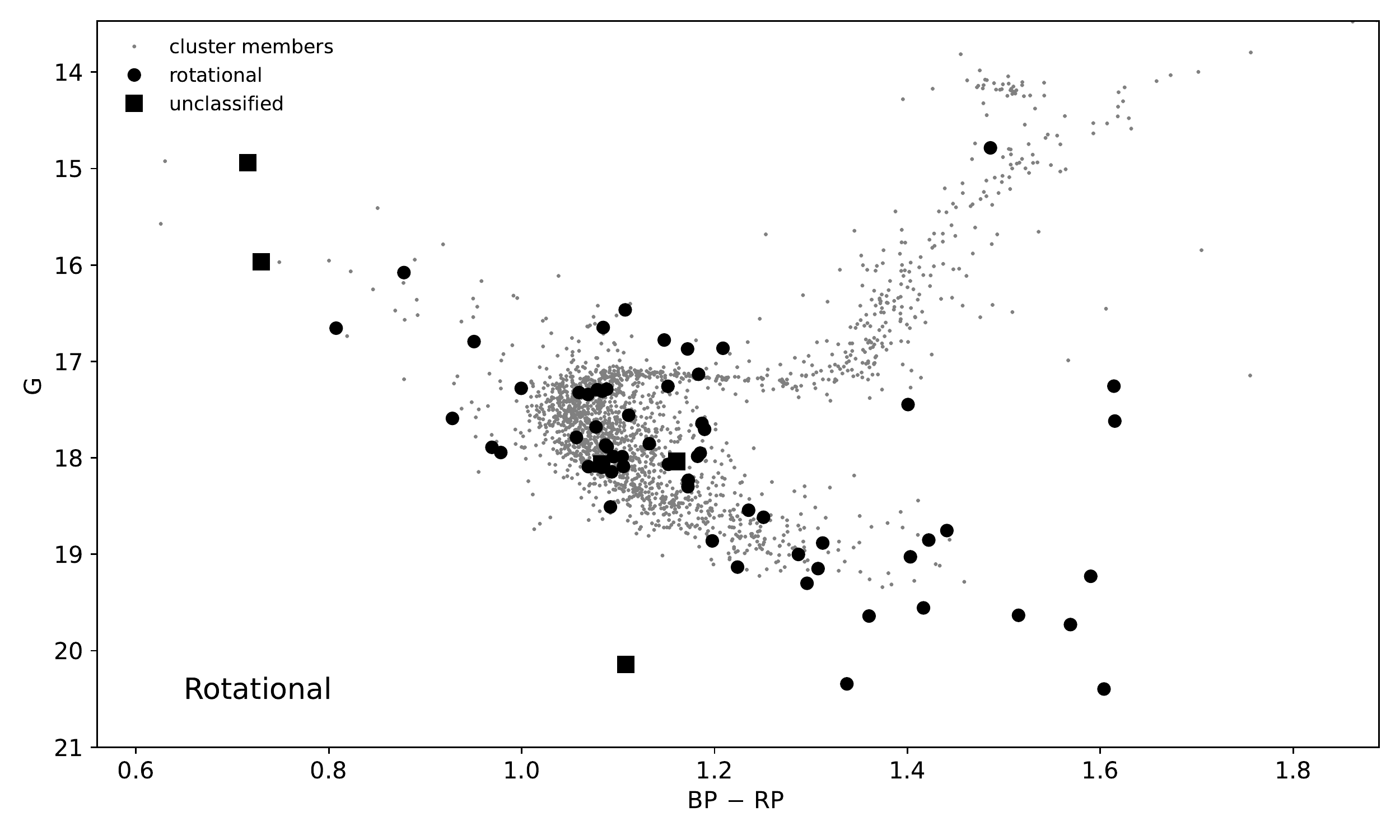}
\caption{The color-magnitude diagram for NGC 6791. We listed variable stars such as binaries, pulsators and rotational cluster members in the top, middle and bottom panels, respectively. Unclassified stars are plotted in the bottom panel.}
\label{fig:3}
\end{figure}

We estimated the mid-times of eclipses or, in the case of non-eclipsing systems, a minimum of a light variation by means of the method described in Kwee \& van\,Woerden\,(1956). We used the mid-times to derive the ephemeris, i.e. a reference epoch, defined as the first mid-time in a given dataset, and an orbital period for each binary system we found. We provided estimates of these two parameters in Table\,1 and Table\,5. The exceptions are systems for which the data are very noisy, not allowing us to derive precise mid-times, and outbursting stars. For these systems we reported only rough estimates (arbitrarily adopted two decimal places for precision) of their orbital periods, derived from the Fourier amplitude spectra. The ephemerides were used to calculate the Observed\,-\,Calculated (O-C) diagrams to check on the orbital period variation. For most of the stars the O-C diagrams do not show any significant period variation. The exceptions are shown in Fig.\,4. \gaia\,EDR3 2051294114497255936 (the top panel of Fig.\,4) shows quite a large amplitude period variation with only one full cycle covered.
These two O-C diagrams are constructed based on an average primary and secondary eclipses. \gaia\,EDR3 2051105342091761536 (the bottom panel of Fig.\,4) shows cyclic variations of the orbital period derived solely from primary eclipses. To explain a periodic change an additional body in the system can be invoked. To test this hypothesis and, if confirmed, to constrain the parameters of those bodies, RV time-series data are required.

\begin{figure}[H]
\centering
\includegraphics[width=0.7\textwidth]{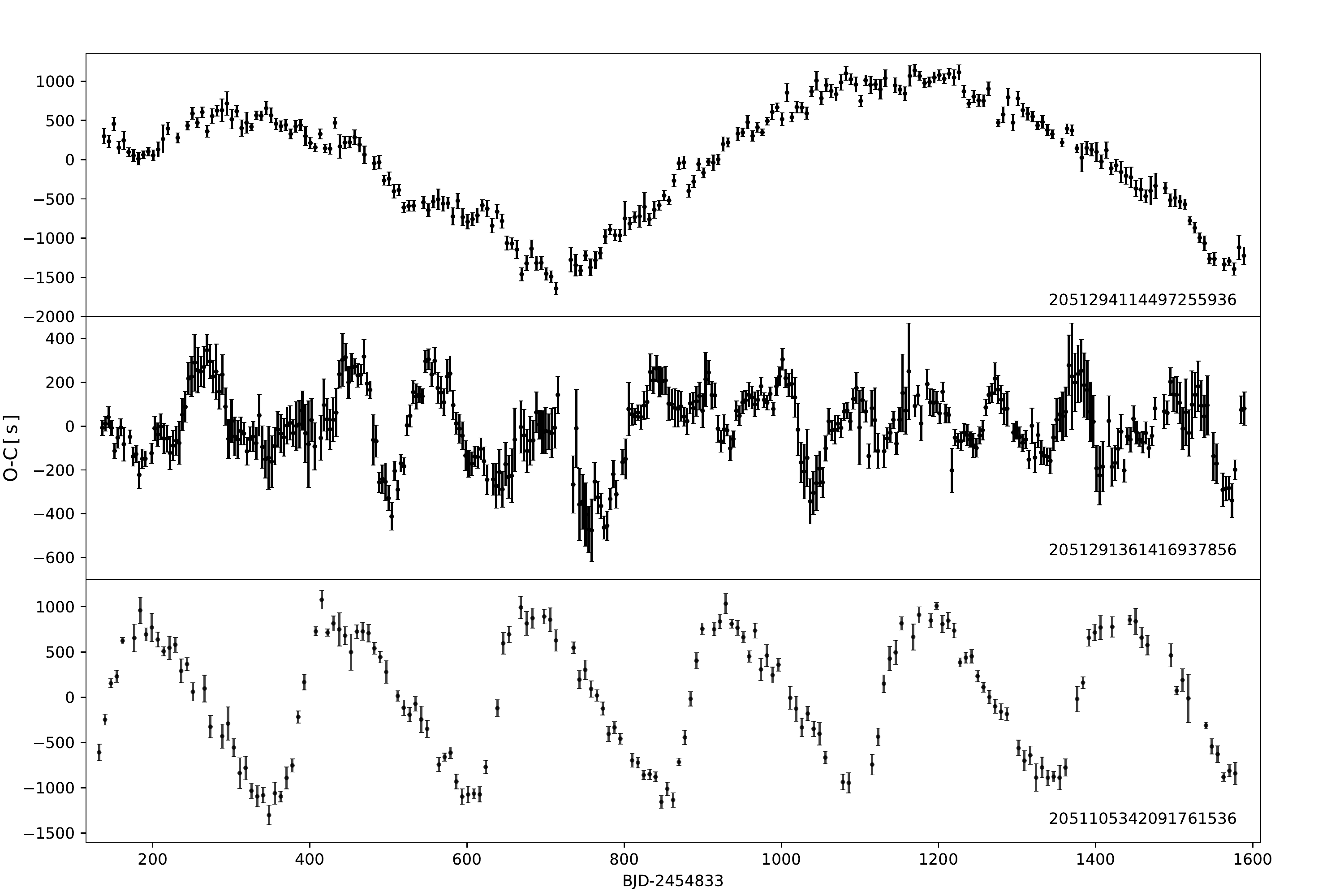}
\caption{O-C diagrams for three binary systems that show variation of the orbital period.}
\label{fig:4}
\end{figure}

\subsection{Pulsators}
We found 70 objects that show an intrinsic variability caused by stellar oscillations. We identified the following types of pulsators, i.e. $\delta$\,Scuti and $\gamma$-Dor, solar-like along the RGB, RR\,Lyr, pulsating hot subdwarf B stars (sdBV) on the EHB, and semi-regular along the RGB and AGB. In Table\,2 we show only the cluster member counterpart, while the field pulsators are listed in Table\,6.

Among the cluster members we found one blue straggler, 36 solar-like pulsators (10 being RC objects, while 26 are still on the RGB), three sdBVs and five semi-regular giants. We analyzed the pulsation component in the solar-like counterpart, and the results will be published elsewhere. The analysis of the only three sdBVs we found in NGC6791 has been already reported by Sanjayan \etl(2022). Among sdBVs only \gaia\,EDR3 2051105509596144768 is a known binary star. The RV is likely influenced by orbital motion. The other two sdBVs also show different RVs from the cluster average, which may be an indication of their binary nature. Sanjayan \etl(2022) reported on the time-series spectroscopy (their Figure\,4), which indicates a RV variation. The RVs of other pulsators we found to be members of NGC\,6791 are close to the average value, with only a few exceptions. For instance, \gaia\,EDR3 2051293255503478528 shows a much faster motion, but the quality flag of the spectra suggests a possible contamination. We do not have resources to sort out these inconsistency cases. Perhaps the explanation of these cases is an additional RV component either of a binary or a rotational nature. In the case of other objects the RV confirms their membership and a single nature. We are aware that their spectra could have been taken while their orbital RV is negligible, but we consider it to be a less likely case. As in the case of binaries, the location of pulsators in the CMD (the middle panels of Fig.\,3 and 6) may still be influenced by binary components.

\begin{table*}
\caption{List of cluster members showing pulsations. See the caption of Table\,1 for explanation. Ref 6 refers to Sanjayan \etl(2022).}
\label{tab:cluster_binaries}
\centering
\renewcommand{\arraystretch}{1.2}
\resizebox{\columnwidth}{!}{%
\begin{tabular}{clccclllll}
\hline\hline
\multicolumn{1}{c}{\multirow{2}{*}{\gaia\,EDR3}} & \multicolumn{1}{c}{\multirow{2}{*}{KIC}} & \multicolumn{1}{c}{G} & \multirow{2}{*}{CMD}& \multirow{2}{*}{HRD} & \multicolumn{1}{c}{T$_{\rm eff}$} & \multicolumn{1}{c}{\multirow{2}{*}{\logg}} & \multicolumn{1}{c}{RV} & \multicolumn{1}{c}{\multirow{2}{*}{[Fe/H]}} & \multicolumn{1}{c}{\multirow{2}{*}{Ref}}\\
&& \multicolumn{1}{c}{[mag]} &&& \multicolumn{1}{c}{[K]} && \multicolumn{1}{c}{[km/s]} &&\\
\hline\hline
\multicolumn{10}{c}{$\gamma$-Dor}\\
2051105479535706752 & 2438249\,$^{17L}$         & 15.573 & BS & BS    & 6\,980(90)	& 3.636(84)	& -56.7(9)	   & -0.15(13)  & 2$^*$ \\
\hline
\multicolumn{10}{c}{solar-like}\\
2051098710667044608 & 2297384\,$^{17L}$         & 14.204 & RC & RC & 4\,512(91)  & 2.460(28)	& -45.71(17)   & 0.400(7)  & 5 \\
2051099260422933632 & 2437539\,$^{17L}$         & 14.545 & RGB & RGB  & 4230(30)  & 2.353(10)	& -39.9(4)   & -0.57(7) & 2$^*$ \\
2051099329142406400 & 2437240\,$^{17L}$         & 14.648 & RGB & RGB  & 4\,380(70)  & 3.116(17)	& -45.5(3)   & -0.12(13) & 2$^*$ \\
{\bf 2051104586182496000} & 2438462\,$^{12L}$         & 14.586 & RGB & RGB  & 4\,240(80)  & 2.020(23)	& -45.71(37)   & -0.39(7)  & 2$^*$  \\
2051105135938204160 & 2437507\,$^{17L}$         & 14.032 & RGB & RGB   & 4\,262(80)  & 2.160(46)	& -47.504(21)  & 0.400(7)  & 5 \\
2051105135938218624 & 2437698\,$^{17L}$         & 14.191 & RC  & RC & 4\,521(93)  & 2.370(28)	& -45.25(6)  & 0.400(8)  & 4 \\
2051105239017444992 & 2437564\,$^{17L}$         & 14.226 & RC  & RC & 4\,515(99)  & 2.400(29)	& -48.53(11)   & 0.400(9)  & 5 \\
2051105342096702080 & 2437804\,$^{17L}$         & 14.110 & RC  & RC & 4\,459(97)  & 2.350(27)	& -45.367(23)  & 0.400(9)  & 5  \\
2051286795872609920 & 2436209\,$^{17L}$         & 14.802 & RGB & RGB & 4\,498(103) & 2.650(45)	& -47.91(45)   & 0.400(9)  & 5 \\
{\bf 2051287002031072768} & 2437164\,$^{12L}$         & 14.157 & RC  & RC &4\,300(50)   & 2.515(10)	& -48.26(37)   & -0.58(13) & 2$^*$ \\
2051287070750544128 & 2436900\,$^{17L}$         & 14.456 & RGB & RGB   & 4\,428(101) & 2.500(45)	& -47.92(8)  & 0.400(9)  & 5 \\
{\bf 2051287242549228032} & 2436608\,$^{10L}$         & 14.446 & RGB & RGB & 4\,500(50)  & 2.254(35)	& -45.5(15)	   & -0.07(7)  & 2$^*$ \\
{\bf 2051287311268703232} & 2436543\,$^{12L}$         & 14.120 & RC  & -- & -- & -- & -- & -- & -- \\
2051287311268706432 & 2436540\,$^{17L}$         & 14.856 & RGB & RGB & 4\,473(104) & 2.560(46)	& -48.13(14)   & 0.400(10)  & 4 \\
2051288067183227008 & 2436417\,$^{17L}$         & 14.104 & RC  & -- & -- & -- & -- & -- & -- \\
2051288135902691840 & 2436332\,$^{17L}$         & 14.302 & RGB & RGB & 4\,360(60)  & 3.445(20)	& -47.54(30)   & -0.07(12) & 2$^*$ \\
2051288170262450816 & 2436458\,$^{17L}$         & 14.532 & RGB & RGB & 4\,250(30)  & 2.602(10)   & -46.36(30)   & -0.24(11) & 2$^*$ \\
2051291228279110912 & 2435987\,$^{17L}$         & 14.529 & RGB & RGB & 4\,428(98)  & 2.490(46)	& -44(4)	   & 0.400(9)  & 5 \\
2051292877546340864 & 2437340\,$^{17L}$         & 13.462 & RGB & RGB & 4\,058(82)  & 1.660(43)	& -45.0(8)   & 0.300(9)  & 5 \\
2051293083704792960 & 2437444\,$^{17L}$         & 13.998 & RGB & RGB & 4\,105(66)  & 2.35(11)	& -56(4)	   & 0.12(6)  & 3 \\
2051293118064539520 & 2437496\,$^{17L}$         & 13.069 & RGB & RGB & 3\,920(67)  & 1.350(45)	& -47.8(7)   & 0.300(7)  & 5 \\
2051293118064542976 & 2437653\,$^{17L}$         & 15.090 & RGB & RGB & 4\,559(97)  & 2.710(48)	& -47(1)	   & 0.400(8)  & 5 \\
{\bf 2051293152424246912} & 2436884\,$^{10L}$         & 13.476 & RGB & RGB & 4\,061(83)  & 1.670(43)	& -44.482(34)  & 0.300(9)  & 4 \\
2051293186784001152 & 2437103\,$^{17L}$         & 14.378 & RGB & RGB & 4\,500(50)  & 3.356(41)	& -48.6(15)   & 0.002(92) & 2$^*$ \\
2051293221143724032 & 2436824\,$^{17L}$         & 14.478 & RGB & -- & -- & -- & -- & -- & -- \\
2051293221143725952 & 2436814\,$^{17L}$         & 14.209 & RGB & -- & -- & -- & -- & -- & -- \\
2051293255503469184 & 2436912\,$^{17L}$         & 14.181 & RC  & RC & 4\,330(90)  & 2.140(46)	& -45.75(42)   & 0.300(8)  & 4 \\
2051293255503478528 & 2437040\,$^{17L}$         & 14.158 & RGB & RGB & 3\,760(50)  & 2.955(27)	& -81.7(8)	   & 0.003(10)  & 5\,$^*$ \\
2051293319920668160 & 2569935\,$^{17L}$         & 13.065 & RGB & RGB & 4\,032(65)  & 1.500(46)	& -47.90(8)  & 0.400(6)  & 5 \\
2051293599100902016 & 2570518\,$^{17L}$         & 14.683 & RGB & RGB & 4\,509(89)  & 2.600(49)	& -48.68(8)	   & 0.300(7)  & 5 \\
2051293977058270592 & 2436732\,$^{17L}$         & 14.135 & RC & RGB & 4\,300(100)  & 2.60(7)  & -44.9(2)   & -0.36(8)  & 2$^*$ \\
2051294045777739520 & 2569360\,$^{17L}$         & 14.093 & RGB & RGB & 4\,287(86)  & 2.180(46)	& -46.44(47)   & 0.300(8)  & 5 \\
2051294114497261952 & 2569618\,$^{17L}$         & 14.797 & RGB & RGB & 4\,486(103) & 2.650(46)	& -45.75(9)	   & 0.400(9)  & 5 \\
{\bf 2051294213273828224} & 2569624\,$^{X}$       & 12.965 & RGB & RGB & 3\,896(62)  & 1.300(45)	& -44.98(7)  & 0.300(6)  & 5 \\
{\bf 2051294251936167424} & 2569204\,$^{11L}$         & 13.293 & RGB & RGB & 3\,967(77)  & 1.430(43)	& -47.909(11)  & 0.300(8)  & 5 \\
2051297275593144192 & 2569055\,$^{17L}$         & 14.173 & RC  & RC & 4\,543(97)  & 2.450(29)	& -46.881(12)  & 0.400(8)  & 5 \\
\hline
\multicolumn{10}{c}{sdBV}\\
2051105509596144768 & 2438324\,$^{35S, 12L}$    & 17.859 & EHB & EHB & 25\,290(300) & 5.510(43) & -90(3)       & -2.62(11) & 6 \\
2051105754408746880 & 2437937\,$^{1S, 1L}$      & 17.878 & EHB & EHB & 24\,860(270) & 5.348(52) & -75(6)       & -2.46(12) & 6 \\
2051294183216739584 & 2569576\,$^{12S, 5L}$     & 17.752 & EHB & EHB & 23\,540(210) & 5.311(35) & -28(4)      & -2.73(23) & 6 \\
\hline
\multicolumn{10}{c}{semi-regular}\\
2051105101578558464 & 2438151\,$^{17L}$         & 15.655 & RGB & MS & 5\,420(60)	& 3.912(19)	& -50.0(15)	   & -0.10(9)  & 1$^*$  \\
2051105616974709504 & 2438421\,$^{17L}$         & 12.246 & AGB & AGB & 3\,555(57)	& 0.510(43)	& -46.540(46)  & 0.200(8)  & 5 \\
2051287002031070208 & 2437171\,$^{17L}$         & 12.394 & AGB & AGB & 3\,540(56)	& 0.610(42)	& -47.11(16)   & 0.300(8)  & 5 \\
2051287173829740032 & 2436324\,$^{8L}$          & 12.373 & AGB & AGB & 3\,332(53)	& 0.270(41)	& -49.315(19)  & 0.200(10) & 5 \\
2051292976324527232 & 2437317\,$^{17L}$         & 14.939 & RGB & -- & -- & -- & -- & -- & -- \\
\hline\hline
\end{tabular}
}
\end{table*}

\subsection{Rotational Variables}
We defined rotational type as stars showing modulated flux variation. Such variation is caused by the presence of migrating spots on the surface, which contribute to a flux modulation as a star rotates. The rotational variables can mimic other types of variability, e.g. ellipsoidal or contact binaries, and high-amplitude radially pulsating stars. We stress that our identification may not be ideal for rotational variable stars. We assumed that binaries show none or negligible modulation of a flux variation. This modulated variation can be verified by either a light curve shape or a profile of peaks in amplitude spectra. A complex profile indicates either period or amplitude change. Objects showing light curves with modulated flux variation have been classified as rotational variables.

We found a total of 145 rotational variables out of which 62 are cluster members and are listed in Table\,3. The field counterpart is listed in Table\,7. Likewise among binaries, there is no \gaia\ astrometry for four rotational variables and they are also listed in Table\,7. We show the CMD location of members in the bottom panel of Fig.\,3. The members occupy mostly the MS region, with three exceptions being the RGB objects and three exceptions being BS objects. Likewise in the case of binaries and pulsators, their true location may be influenced by binarity.

\begin{table*}
\caption{List of cluster members showing rotational variability. See the caption of Table\,1 for explanation.}
\label{tab:cluster_rotational}
\centering
\resizebox{\columnwidth}{!}{%
\begin{tabular}{clrccclllll}
\hline\hline
\multicolumn{1}{c}{\multirow{2}{*}{\gaia\,EDR3}} & \multicolumn{1}{c}{\multirow{2}{*}{KIC}} & \multicolumn{1}{c}{Period} & \multicolumn{1}{c}{G} & \multirow{2}{*}{CMD} & \multirow{2}{*}{HRD} & \multicolumn{1}{c}{T$_{\rm eff}$} & \multicolumn{1}{c}{\multirow{2}{*}{\logg}} & \multicolumn{1}{c}{RV} & \multicolumn{1}{c}{\multirow{2}{*}{[Fe/H]}} & \multicolumn{1}{c}{\multirow{2}{*}{Ref}}\\
&& \multicolumn{1}{c}{[days]} & \multicolumn{1}{c}{[mag]} &&& \multicolumn{1}{c}{[K]} && \multicolumn{1}{c}{[km/s]} &&\\
\hline\hline
2051099015605500672 & 2297584\,$^{X}$      & 18.02     & 18.299 & MS$^+$ & -- & -- & -- & -- & -- & -- \\
2051104757981275264 & 2438685\,$^{X}$    & 7.78      & 17.256 & RGB & MS &5\,500(50)	& 4.953(20)	& -44.8(44)	 & 0.47(15)  & 1\,$^{*}$ \\
{\bf 2051104891118311040} & 2437994\,$^{X}$       & 1.85     & 19.228 & MS$^+$ & -- & -- & -- & -- & -- \\
{\bf 2051104891120796672} & 2437990\,$^{X}$      & 13.15     & 19.148 & MS$^+$ & -- & -- & -- & -- & -- & -- \\
2051104959840285696 & 2437791\,$^{X}$      & 15.28     & 18.067 & MS$^+$ & -- & -- & -- & -- & -- & -- \\
{\bf 2051104964139549824} & 2437941\,$^{X}$      & 10.12     & 17.944 & MS & -- & --           &    --        & --          & --              & -- \\
2051104994200024448 & 2438113\,$^{X}$      & 5.80    & 17.288 & MS & -- &  -- & -- & -- & -- & -- \\
2051105097281783040 & 2438031\,$^{X}$      & 8.28    & 17.591 & MS & MS & 5\,810(40)	& 4.216(34)	& -62.3(23)	 & -0.158(47)  & 1\,$^{*}$ \\
2051105131636983808 & 2437646\,$^{X}$      & 13.23     & 17.558 & MS & -- & -- & -- & -- & -- & --\\
2051105234715895552 & 2437707\,$^{X}$      & 11.00     & 17.867 & MS & MS & 6\,080(30)	 & 4.641(35)  & -47.8(48)	  & 0.60(14)  & 1\,$^{*}$ \\
2051105273377206784 & 2437801\,$^{X}$      & 7.19      & 18.508 & MS$^+$ & MS & 5\,380(70)	 & 4.111(26)  & -65.30(10)	& -0.40(10)& 1\,$^{*}$ \\
{\bf 2051105342091728384} & 2437761\,$^{X}$      & 11.81     & 19.001   & MS & -- & -- & -- & -- & -- & -- \\
2051105372154852736 & 2437849\,$^{4L}$     & 11.02     & 16.863 & MS & MS & 5\,260(50)	& 3.923(16)	& -51.7(14)	 & -0.31(11)   & 1\,$^{*}$ \\
2051105372154855552 & 2437944\,$^{X}$      & 1.96      & 18.543 & MS$^+$ & -- & -- & -- & -- & -- & -- \\
{\bf 2051105372157195264} & --                     & 2.45      & 19.556 & MS$^+$ & MS & 5480(50)	& 4.835(21)	& -56(4)	 & 0.45(17)   & 1\,$^{*}$ \\
2051105578318626304 & 2438569\,$^{X}$      & 7.00      & 17.279 & MS & MS & 5\,970(50)	& 4.295(20)	& -69.2(15)	 & 0.32(14)  & 1\,$^{*}$ \\
{\bf 2051105612675379968} & 2438390\,$^{X}$      & 3.80      & 17.987 & MS & -- & -- & -- & -- & -- & -- \\
{\bf 2051105685689315968} & 2438129\,$^{X}$      & 16.58     & 17.853 & MS & -- & -- & -- & -- & --& -- \\
2051105784484742144 & 2570443\,$^{X}$      & 2.20      & 17.643 & MS & MS & 5\,610(50)	& 4.298(20)	& -133.2(45)	 & -0.62(30)  & 1\,$^{*}$ \\
{\bf 2051105818833811584} & 2438344\,$^{X}$      & 7.74      & 18.616 & MS$^+$ & -- & -- & -- & -- & -- & -- \\
2051105857492894592 & 2570649\,$^{X}$      & 13.36     & 17.293 & MS & -- & -- & -- & -- & -- & -- \\
{\bf 2051105921913036160} & 2570622\,$^{X}$      & 9.57      & 19.026  & MS$^+$ & -- & -- & -- & -- & -- & --\\
{\bf 2051105926212896512} & 2570559\,$^{X}$      & 1.32      & 20.343 & MS$^+$ & -- & -- & -- & -- & -- & -- \\
{\bf 2051107055787471104} & 2438631\,$^{X}$      & 7.36      & 19.729 & MS$^+$ & -- & -- & -- & -- & -- & -- \\
{\bf 2051287002024520064} & --                     & 10.86     & 19.640 & MS$^+$ & -- & -- & -- & -- & -- \\
{\bf 2051287066449451136} & 2436942\,$^{X}$      & 9.26      & 18.098 & MS & -- & -- & -- & -- & -- & -- \\
2051287066449464064 & 2436959\,$^{X}$      & 2.65      & 18.852 & MS$^+$ & -- & -- & -- & -- & -- & -- \\
2051287070744003456 & 2436969\,$^{X}$      & 7.92      & 19.132 & MS$^+$ & RGB & 4\,500(50)	 & 3.09(8)  & -45.0(8)  & -0.155(20)  & 2\,$^{*}$ \\
2051287242549232384 & 2436767\,$^{X}$      & 12.32     & 17.309 & MS & MS & 5\,690(50)	 & 4.163(20)  & -43.2(57)	  & 0.148(33)  & 1\,$^{*}$ \\
{\bf 2051287895379273088} & 2436011\,$^{X}$      & 10.00     & 19.632 & MS$^+$ & -- & -- & -- & -- & -- & -- \\
2051288372119657728 & 2569185\,$^{X}$      & 13.43     & 17.258 & MS & MS & 5\,660(50)	 & 4.117(34)  &	-97.9(35)  &	-0.08(10) & 1\,$^{*}$ \\
2051288376420896384 & 2569162\,$^{X}$      & 5.90      & 18.088 & MS & -- & -- & -- & -- & -- & -- \\
2051291262638872192 & 2568864\,$^{X}$      & 14.42    & 17.618 & MS & MS & 5\,370(30)	& 4.78(10)	& -61.5(42)	 & 0.58(27)   & 1\,$^{*}$ \\
{\bf 2051291503157017472} & 2568685\,$^{X}$      & 19.87     & 17.342 & MS & -- & -- & -- & -- & -- & -- \\
2051292873245240192 & 2437350\,$^{X}$      & 16.82     & 17.885 & MS & MS & 5\,790(50)	 & 4.913(20)  & -74.3(15)  & 0.13(8)  & 1\,$^{*}$ \\
{\bf 2051292911906089728} & 2437521\,$^{X}$      & 11.27     & 17.987 & MS$^+$ & -- & -- & -- & -- & -- & -- \\
{\bf 2051292946259301504} & 2437092\,$^{2L}$     & 3.20      & 18.145 & MS$^+$ & MS & 4\,700(100)	 & 4.619(50)  & -78(1) & -1.87(6) & 1\,$^{*}$ \\
2051293014978876672 & 2437584\,$^{X}$      & 5.90      & 18.486 & --$^+$ & -- &  -- & -- & -- & -- & -- \\
2051293083704791040 & 2437354\,$^{8L}$     & 7.21      & 16.465 & MS & MS & 5\,580(50)	 & 4.391(20)  & -55.6(8)&	-0.413(46) & 2\,$^{*}$ \\
2051293113776412416 & 2437613\,$^{X}$      & 11.05     & 17.787 & MS & -- &  -- & -- & -- & -- & -- \\
2051293148133456896 & 2437062\,$^{X}$      & 8.82      & 17.133 & RGB & -- & -- & -- & -- &  -- & -- \\
2051293152424243968 & 2436909\,$^{X}$      & 12.00     & 16.778 & MS & MS & 5\,640(40)	 & 4.654(21)  & -69.3(26)	  & -0.203(17)  & 1\,$^{*}$ \\
2051293186783997056 &  --                    & 9.63      & 18.232 & MS$^+$ & -- & -- & -- & -- & -- & -- \\
{\bf 2051293186784004096} & 2437238\,$^{X}$    & 4.00     & 16.079 & BS & BS & 5\,880(30)	 & 3.546(23)  & -48.0(4)&	-0.34(8) & 2\,$^{*}$ \\
{\bf 2051293255503480320} & 2437079\,$^{X}$      & 5.81      & 16.794 & BS & -- & -- & -- & -- & -- & -- \\
{\bf 2051293289863230720} & 2437338\,$^{X}$      & 1.24      & 16.655 & BS & -- & -- & -- & -- & -- & -- \\
2051293319934813184 & 2569984\,$^{X}$      & 12.50     & 17.704 & MS & -- & -- & -- & -- & -- & -- \\
2051293358583014528 & 2569763\,$^{4L}$     & 8.29      & 16.648 & MS & MS/RGB & 5\,510(20)	 & 3.788(49)  & -38.6(5)  & -0.397(48) & 2\,$^{*}$ \\
2051293392942762240 & 2569825\,$^{X}$      & 0.48      & 18.754 & MS$^+$ & MS & 5\,000(50)	& 4.86(15)	& -86.3(7)    &	0.02(14) & 1\,$^{*}$ \\
2051293427302195584 & 2437884\,$^{X}$      & 3.23      & 16.871 & MS$^+$ & MS & 5\,930(20)	& 4.048(26)	& -43.9(3) & -0.44(10)  & 2\,$^{*}$ \\
2051293530375009920 & 2570217\,$^{X}$      & 5.55      & 17.891 & MS & MS & 5\,760(20)	& 4.49(7)	& -62.4(23)     &	0.07(8)  & 1\,$^{*}$ \\
2051293663519464192 & 2570420\,$^{X}$      & 19.96     & 17.951 & MS & -- & -- & -- & -- & -- & -- \\
2051293977053182592 & 2436790\,$^{X}$      & 9.61      & 18.861 & MS & MS & 4\,990(80)	 & 4.832(25)  & -169(3)	  & -0.356(15)  & 1\,$^{*}$ \\
{\bf 2051294007116759808} & 2569597\,$^{X}$      & 7.17      & 18.091 & MS$^+$ & -- & -- & -- & -- & -- & -- \\
{\bf 2051294148851985536} & --                     & 9.66      & 20.397 & MS$^+$& -- & -- & -- & -- & -- & -- \\
{\bf 2051294148857020160} & 2569767\,$^{X}$      & 2.40      & 17.323 & MS$^+$ & -- & -- & -- & -- & -- & -- \\
2051294217574157440 &  --                    & 4.92      & 19.300 & MS$^+$ & -- & -- & -- & -- & -- & -- \\
{\bf 2051294251931077888}  & 2569279\,$^{X}$      & 1.14      & 18.883 & MS$^+$& -- & -- & -- & -- & -- & -- \\
{\bf 2051294281994746880} & 2569324\,$^{X}$      & 10.90     & 17.983 & MS  & -- & -- & -- & -- & -- & -- \\
2051294286296294528 & 2569334\,$^{X}$      & 6.29      & 17.447 & MS & MS & 5\,330(20)	 & 4.220(25)  & -47.4(23)  & 0.15(8) & 1\,$^{*}$ \\
2051294797390943872 & 2569591\,$^{X}$      & 14.89     & 17.679 & MS & -- & -- & -- & -- & -- & -- \\
{\bf 2051296176081696768} & 2570281\,$^{12L}$    & 14.42     & 14.785 & RGB & -- & -- & -- & -- & -- & -- \\

\hline\hline
\end{tabular}
}
\end{table*}

\subsection{Unclassified and unidentified variables}
In the case of nine objects, we were unable to unambiguously classify them according to the three types described above. The amplitude of a flux variation is low and it is not clear if the variation remains stable over time. The latter indicates that it is not unlikely these stars may be rotational variables. We leave a definite classification for further analysis, preferentially if based on better quality data. Five members of the cluster are listed in Table\,4 and they are plotted in the bottom panel of Fig.\,3, while four non-members are listed in Table\,8.

While checking the pixel content we found signals in the amplitude spectra, associated with optical detections found in Pan-STARRS data (Chambers \etl 2016, Flewelling \etl 2020) survey that do not have any designations. We classified the signal to a proper variability type and estimated its period. We show the list of these unidentified objects in Table\,9. Since the stars are not listed in the \gaia\ catalog, we are unable to estimate their membership. At three locations on the silicons (last three coordinates listed), we detected a signal in the amplitude spectra, however, we are unable to associate these coordinates with any optical objects in the Pan-STARRS survey. In addition, the last two signals listed may be of the same origin, though the coordinates are different. This result indicates that these three signals listed are residual signals of some other variable stars in the \kep\ field of view that were spread over the silicons. We consider the signals detected at those three specific locations on the detector to be artifacts and not real sources.

\begin{table*}
\caption{List of cluster members with unclassified variability. See the caption of Table\,1 for explanation.}
\label{tab:cluster_unclassified}
\centering
\resizebox{\columnwidth}{!}{
\begin{tabular}{clllccc}
\hline\hline
\multicolumn{1}{c}{\multirow{2}{*}{\gaia\,EDR3}} & \multicolumn{1}{c}{\multirow{2}{*}{KIC}} & \multicolumn{1}{c}{Period} & \multicolumn{1}{c}{G} & \multirow{2}{*}{CMD} &  \multirow{2}{*}{HRD}\\
&& \multicolumn{1}{c}{[days]} & \multicolumn{1}{c}{[mag]} &&\\
\hline\hline
2051105269075634176 & 2437760\,$^{X}$ & 0.44 & 18.064 & MS & --\\
{\bf 2051288067182949760} & 2436421\,$^{8L}$ & 1.44 & 14.937 & BS & --\\
{\bf 2051291498855854208} & 2568724\,$^{X}$ & 0.63 & 20.141 & MS$^+$ & --\\
{\bf 2051293049345064832} & 2437745\,$^{X}$ & 1.43 & 15.968 & BS & --\\
{\bf 2051294148857016192} & 2569676\,$^{X}$ & 0.27 & 18.037 & MS$^+$ & --\\
\hline\hline
\end{tabular}
}
\flushleft
\end{table*}

\section{The distance and age estimation}
We downloaded a grid of isochrones given in the \gaia\ photometric system from the MESA Isochrones and Stellar Tracks (MIST) project (Choi \etl 2016, Dotter 2016). The current version of MIST is 1.2. The MESA version 7503 was employed to calculate isochrones. We selected V/V$_{\rm crit}$\,=\,0. The grid covers age in a logarithmic scale between 9.8 and 10.3 with a step of 0.01, while [Fe/H] was between 0.20 and 0.45 with a step of 0.01.

At least 50$\%$ of the stars are expected to be in binary systems. The observed magnitude of a given target may include the flux contribution from all companions and not a single star, which shifts the position of the target in the CMD. Therefore, we excluded outlying stars where their positions in the CMD are uncertain. For the fit, we only kept the MS, RGB and RC stars. We included magnitude uncertainties as weights in the fit, which prevented the MS targets from over-fitting. The RC and RGB targets, even though less numerous, are brighter, and hence also remains significant in the fit.

The MIST synthetic isochrones are given in absolute magnitudes, and we selected no extinction. To account for a distance and a non-vacuum environment, in our fit, we included a shift (m-M) in the \gaia\ G magnitude and in B$_{\rm p}$-R$_{\rm p}$ color.
The best-fit isochrones point to a narrow range of age and [Fe/H]. The age is 8.91\,Gyr, while [Fe/H] is between 0.26 and 0.28. The apparent distance modulus\,(m-M) equals 13.424, while E(B$_{\rm p}$-R$_{\rm p}$) from  0.165 to 0.176. We show the fits in Fig.\,5. Based on the extinction curve from Cardelli \etl(1989), Bressan \etl(2012) showed a rough relation between extinction in the \gaia\ G band and E(B$_{\rm p}$-R$_{\rm p}$), which is A$_{\rm G} \approx $2$\cdot$E(B$_{\rm p}$-R$_{\rm p}$). Taking the average of E(B$_{\rm p}$-R$_{\rm p}$)\,=\,0.171 and using this relation, we derived A$_{\rm G}$\,=\,0.342\,mag. Subtracting A$_{\rm G}$ from (m-M) we find the true distance modulus of 13.082, which gives the distance to the cluster of 4134\,pc. We also derived distance from the parallaxes of the cluster members with probability membership at least 90\% and the relative parallax uncertainty of smaller than 10\%, which equals 4123(31)\,pc. The distance estimated from the isochrone fit lies well within the uncertainty of the distance estimations from the cluster parallax. The cluster age we derived from the isochrone fit is also comparable to the age reported by Choi \etl(2018).

To verify the correctness of the spectroscopic fits we also engaged isochrones in T$_{\rm eff}$ and \logg\ plane (HRD). We used the isochrones for the age and the average metallicity, which we derived from the isochrone fitting in the CMD. Then, we overplotted the isochrones with our variable stars from Tables\,1-3, for which T$_{\rm eff}$ and \logg\ are listed. The column 'HRD' in Tables\,1-3 describes the location of a given star in the HRD. If the location agrees with the one in CMD, we can expect the spectroscopic fit is likely correct and we obtained an agreement for the majority of stars. It should be noted that the scatter in the HRD is larger, hence some objects may have not been allocated properly. There are clear exceptions, eight among binaries, two among pulsators and two among rotational variables. These cases should be treated with caution.

\begin{figure}[H]
\centering
\includegraphics[width=0.8\textwidth]{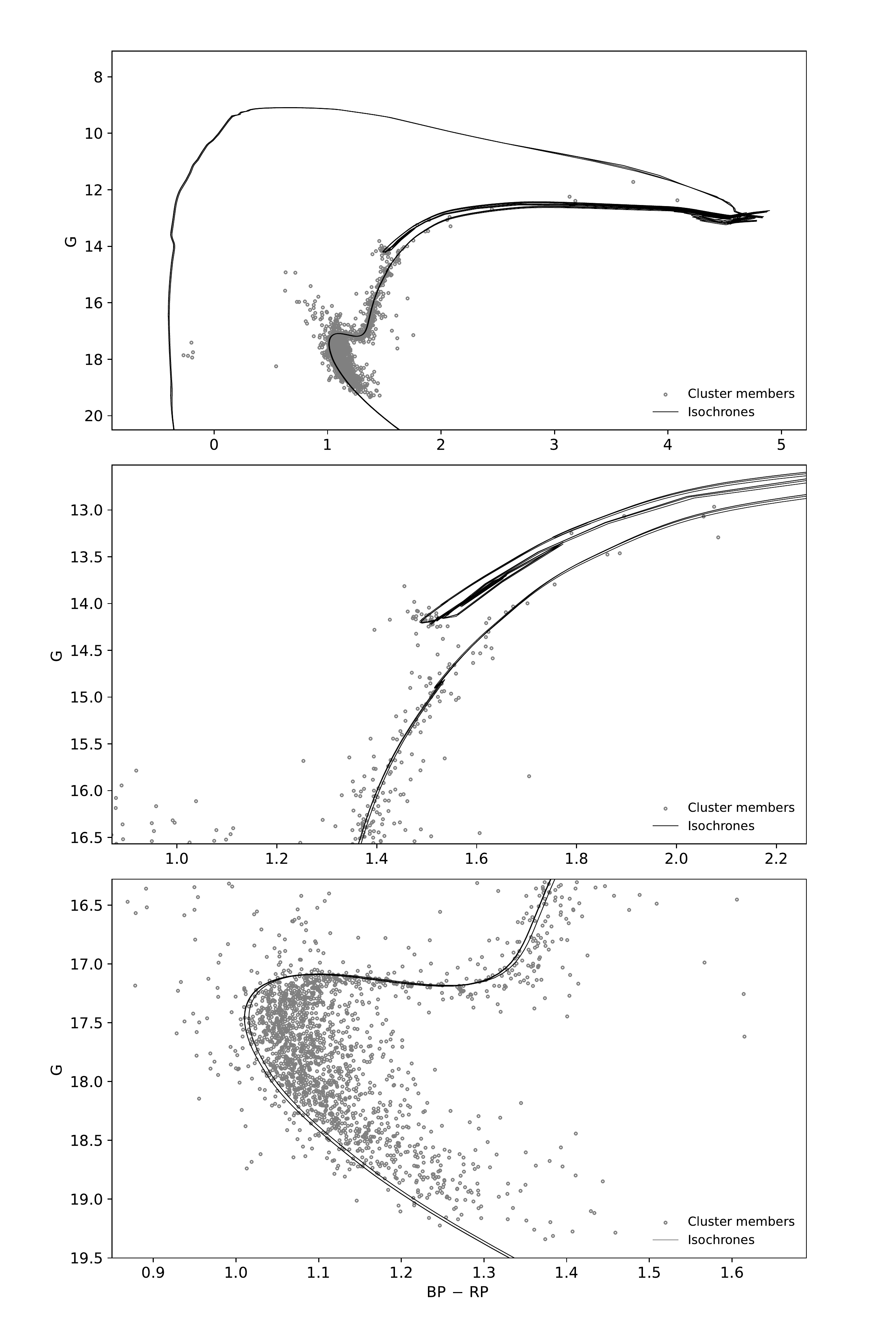}
\caption{The color-magnitude diagram of NGC\,6791, showing the best MIST isochrone fits. The top panel shows the overall diagram, while the middle panel shows the red clump region and the bottom panel shows the main-sequence region.}
\label{fig:5}
\end{figure}

\begin{figure}[H]
\centering
\includegraphics[width=0.7\textwidth]{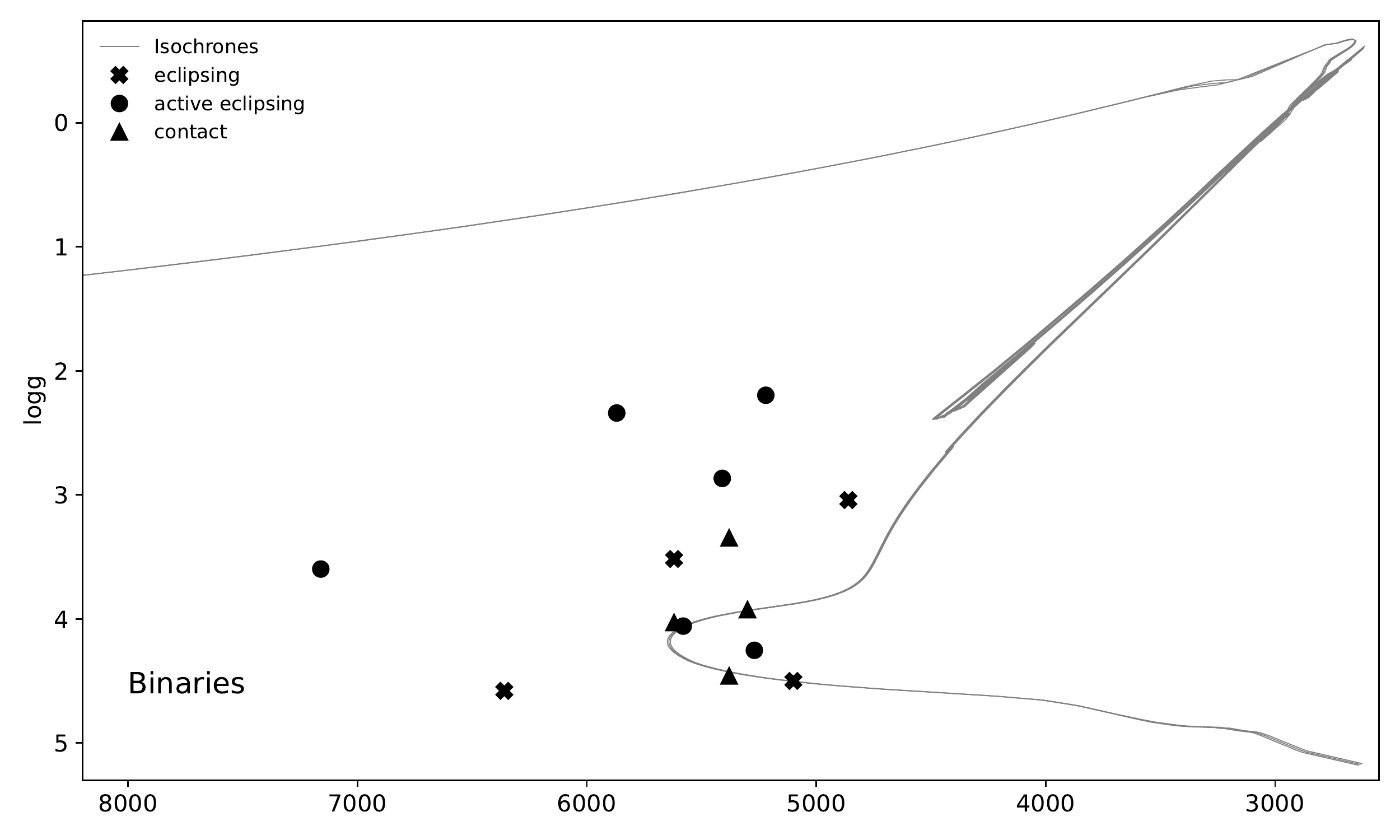}
\includegraphics[width=0.7\textwidth]{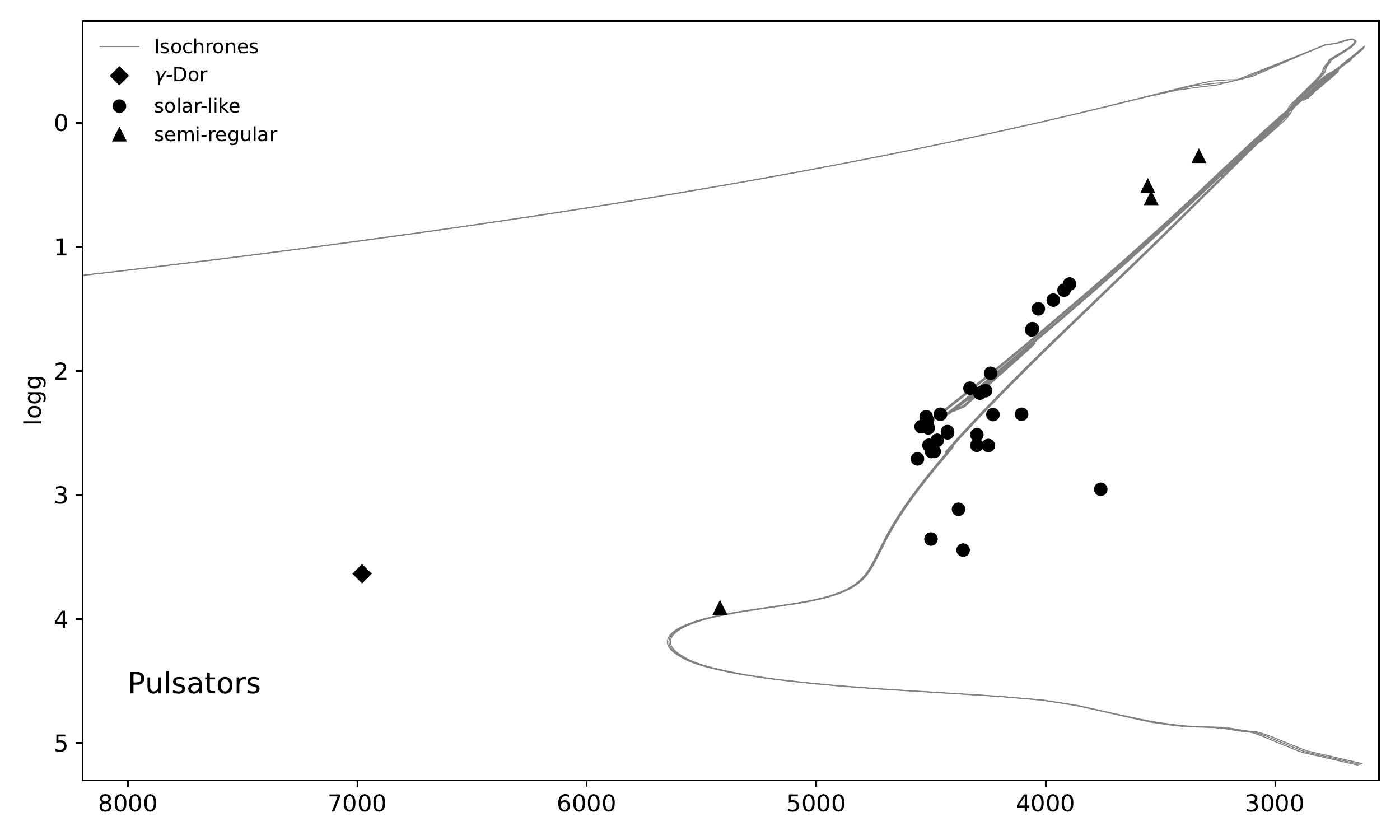}
\includegraphics[width=0.7\textwidth]{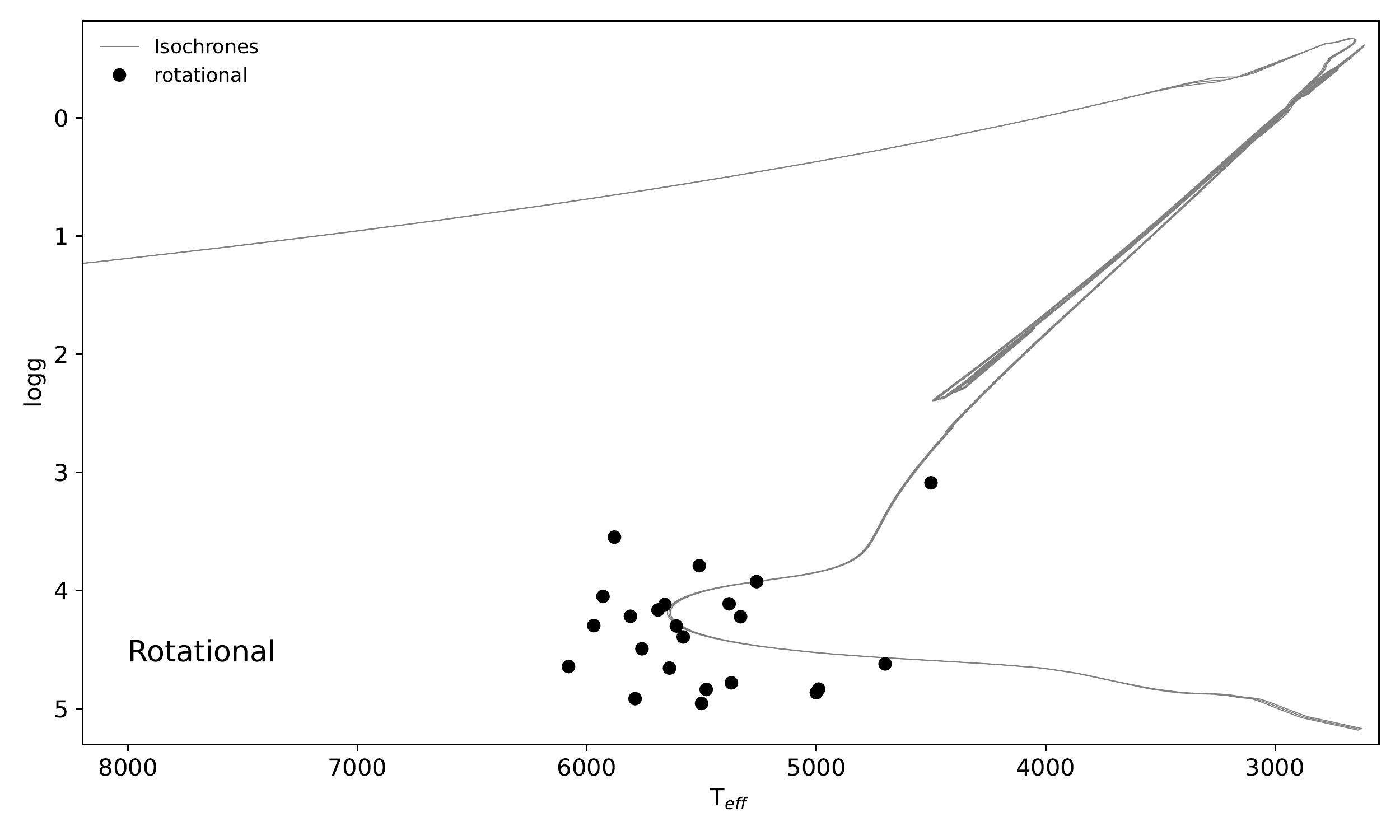}
\caption{The T$_{\rm eff}$--\logg\ diagram showing the cluster members with atmospheric parameters derived in this work. The top, middle and bottom panels show binary, pulsating and rotational stars, respectively. The isochrones represent the best MIST fits in T$_{\rm eff}$ and \logg\ coordinates.}
\label{fig:6}
\end{figure}

\begin{table*}
\caption{List of binary stars that are not cluster members. See the caption of Table\,1 for explanation. Targets with no astrometry available, hence with no membership established, are denoted in {\it italic}.}
\label{tab:field_binaries}
\centering
\resizebox{\columnwidth}{!}{
\begin{tabular}{clllclllll}
\hline\hline
\multicolumn{1}{c}{\multirow{2}{*}{\gaia\,EDR3}} & \multicolumn{1}{c}{\multirow{2}{*}{KIC}} & \multicolumn{1}{c}{P$_{\rm orb}$} & \multicolumn{1}{c}{T$_0$} & \multicolumn{1}{c}{G} & \multicolumn{1}{c}{T$_{\rm eff}$} & \multicolumn{1}{c}{\multirow{2}{*}{\logg}} & \multicolumn{1}{c}{RV} & \multicolumn{1}{c}{\multirow{2}{*}{[Fe/H]}} & \multicolumn{1}{c}{\multirow{2}{*}{Ref}}\\
&& \multicolumn{1}{c}{[days]} & \multicolumn{1}{c}{[BJD]} & \multicolumn{1}{c}{[mag]} & \multicolumn{1}{c}{[K]} && \multicolumn{1}{c}{[km/s]} &&\\
\hline\hline
\multicolumn{10}{c}{eclipsing}\\
{\bf 2051104826700758016} & 2438661\,$^{X}$      & 196.34051(1)    & 2\,455\,037.21466(43)    & 16.449 &    --       &    --       &  --         &    --           & --\\
2051105131638986752   &           --           & 23.878352(36)   & 2\,454\,967.0873(12)     & 19.782    &  --         &  --         &  --         &  --             & --\\
2051107128803742336   & 2438562\,$^{X}$      & 10.406512(6)    & 2\,454\,965.58747(47)    & 19.369    & 6\,220(130)	& 2.864(51) & -49(4)	& -1.29(17) & 2$^*$ \\
2051287311268711168   & 2436579\,$^{X}$      & 9.287153(22)    & 2\,454\,971.1134(18)     & 19.613    &    --       &    --       &  --         &    --           & --\\
2051291739386261248   & 2568780\,$^{X}$      & 3.5316020(11)   & 2\,454\,964.71245(25)    & 19.913    & --	& --	& --	& --  & -- \\
2051292907615354240   & 2437452\,$^{1S,17L}$& 14.4699358(24)  & 2\,454\,974.81692(14)    & 17.286     & 6\,170(50)	& 4.26(9)	& -92.1(6)& -0.86(9) & 2$^*$ \\
2051293079414230912   & 2437505\,$^{10L}$    & 21.476420(11)   & 2\,454\,982.26244(40)    & 18.324    & 4\,920(50)	& 3.25(6)	& -19.6(30) 	    & -0.25(19)  & 1$^*$ \\
\textit{\textbf{2051293118063233408}} & 2437675\,$^{X}$      & 1.75          &  --        & 21.175     & --          &    --       &  --         &    --           & --\\
{\bf 2051294629893394176} & 2569880\,$^{X}$      & 0.51323254(22)  & 2\,454\,964.60382(35)    & 17.866 & --          &    --       &  --         &    --           & --\\
\hline
\multicolumn{10}{c}{active eclipsing}\\
2051104586182512128   & 2438502\,$^{6S,16L}$& 8.35905(26)     & 2\,455\,005.2748(14)     & 16.216     & 5\,563(234)	& 4.17(37)  & -14(5)	  & 0.37(22)    & 3\\
2051104684959920896   & 2438464\,$^{X}$      & 0.43657907(7)   & 2\,454\,964.92853(13)    & 19.443    & 4\,310(50)	& 4.169(20) & 41(16) & -0.91(6)   & 1$^*$ \\
{\bf 2051286761512868992}  & 2436203\,$^{X}$      & 2.06          &     --                 & 14.328    &    --       &  --         &    --      &  --              &-- \\
2051288170257319808  &      --                & 3.7024382(14)   & 2\,454\,968.16622(31)    & 20.135   & --	& --   & --	   & --  & -- \\
2051288372119654528   & 2569138\,$^{X}$      & 0.88305272(12)  & 2\,454\,964.71084(11)    & 19.205    & --	& --   & --	   & --  & -- \\
2051293599094508032   &    --                  & 2.8516350(19)   & 2\,454\,967.1274(6)      & 20.335  & --	& --   & --	   & --  & -- \\
\hline
\multicolumn{10}{c}{contact}\\
2051105543955885312   & 2438471\,$^{X}$      & 0.27345629(34)  & 2\,454\,964.8195(11)     & 19.527    & 4\,300(50)& 3.92(7)	    & 3.9(9)	& -1.354(25) & 1$^*$ \\
2051288269040355584   & 2436044\,$^{X}$      & 0.24533697(21)  & 2\,454\,964.6646(7)      & 19.452    &   --        &     --      &      --     &        --       & --\\
{\bf 2051286417915483392} & 2297170\,$^{X}$      & 0.3664306(6)    & 2\,455\,002.6182(13)     & 17.797 &    --       &      --     &     --      &        --       & --\\
2051287070750548736   & 2437038\,$^{17L}$    & 0.267678275(46) & 2\,454\,964.58218(14)    & 16.149    &  --         &    --       &    --       &     --          & --\\
2051291228279113216   & 2435971\,$^{17L}$    & 0.27182769(8)   & 2\,454\,964.81810(25)    & 16.276    &   --        &   --        &   --        &    --           & --\\
2051291468797350528   & 2568971\,$^{17L}$    & 5.088522(10) & 2454966.2350(17) & 13.081          &    --       &     --      &     --      &    --           & --\\
2051293530381417856   & 2570289\,$^{17L}$    & 0.279027932(31) & 2\,454\,964.57880(1)     & 15.781    &  --         &   --        &   --        &    --           & --\\
{\bf 2051295076563264000}  & 2570552\,$^{X}$      & 0.28417822(44)  & 2\,454\,964.5913(13)     & 20.279&  --         &   --        &  --         &    --           & --\\
2051295454527201792   & 2570460\,$^{X}$      & 0.2659879(15)   & 2\,454\,964.61531(47)    & 18.689    &    --       &   --        &     --      &  --             & --\\
2051296313520692864   & 2708123\,$^{X}$      & 0.302184513(81) & 2\,454\,964.69239(22)    & 17.742    &  --         &   --        &    --       &    --           & --\\
2051297374371300992   & 2569082\,$^{X}$      & 0.33401898(23)  & 2\,454\,964.8252(6)      & 17.579    & 5\,750(50) & 4.052(15)	 & -38.30(54)	   & -1.069(20)          & 1$^*$ \\
\textit{\textbf{2051297477460338304}} &  -- & 0.2690446(1)    & 2\,454\,964.7175(30)     & 20.838        &    --       &     --      &     --      &    --           & --\\
\hline
\multicolumn{10}{c}{outbursting}\\
2051287203888388736   & 2436450\,$^{15S,12L}$ & 35.71         &        --            & 19.884    & --          &   --        &     --      &    --           & --\\
\hline\hline
\end{tabular}
}
\end{table*}

\begin{table*}
\caption{List of pulsators that are not cluster members. See the caption of Table\,1 for explanation.}
\label{tab:Pulsators_field}
\centering
\resizebox{\columnwidth}{!}{%
\begin{tabular}{clclllll}
\hline\hline
\multicolumn{1}{c}{\multirow{2}{*}{\gaia\,EDR3}} & \multicolumn{1}{c}{\multirow{2}{*}{KIC}} & \multicolumn{1}{c}{G} & \multicolumn{1}{c}{T$_{\rm eff}$} & \multicolumn{1}{c}{\multirow{2}{*}{\logg}} & \multicolumn{1}{c}{RV} & \multicolumn{1}{c}{\multirow{2}{*}{[Fe/H]}} & \multicolumn{1}{c}{\multirow{2}{*}{Ref}}\\
&& \multicolumn{1}{c}{[mag]} & \multicolumn{1}{c}{[K]} && \multicolumn{1}{c}{[km/s]} &&\\
\hline\hline
\multicolumn{8}{c}{$\delta$-scuti}\\
2051098981244205056  & 2297728\,$^{1S,18L}$  & \multicolumn{1}{r}{9.929}       & 6\,888(25)  &	3.680(40)& -37(7)      & 0.272(22)     & 3 \\  
2051107369321426432 & 2570760\,$^{18L}$      & 13.459     &  --        &  --        &    --         &         --      & --\\
\hline
\multicolumn{8}{c}{solar-like}\\
{\bf 2051098710667526912} & 2297357\,$^{1L}$       & 12.115      &    --      &   --        &    --         &     --          & --   \\
{\bf 2051099054264914048} & 2437622\,$^{12L}$      & 15.583      &    --      &   --        &   --          &     --          &  --  \\
{\bf 2051099329142404480} & 2437207\,$^{X}$        & 18.731      &    --      &    --       &   --          &     --          &   -- \\
2051104105146054784 & 2297793\,$^{17L}$      & 13.626      & 4\,085(81) & 1.430(52) & -83.615(30) & -0.300(10)    & 5  \\
2051104139506469760 & 2438094\,$^{17L}$      & 15.322      & 4\,060(50)   & 1.11(15)    & -84.7(6)      & -1.14(8) & 2$^*$ \\
2051104861060304640 & 2437816\,$^{17L}$      & 13.958      & 4\,229(77) & 2.090(47) & 0.400(7)    & -47.27(9)     & 5  \\
{\bf 2051105720053903232} & 2438289\,$^{10L}$      & 14.384      &  --        &   --        &     --        &   --            &  --  \\
2051105857492905600 & 2570715\,$^{7L}$       & 12.678      & 4\,713(84) & 2.330(36) & -3.30(7)    & -0.200(7)     & 5  \\
{\bf 2051107369321441920} & 2570794\,$^{10L}$      & 15.158      &    --      &     --      &     --        &    --           &   -- \\
2051286520994706432 & 2436457\,$^{18L}$      & 13.245      & 4\,969(24) & 3.015(39) & -19.14(4)   & -0.422(22)    & 3  \\
{\bf 2051286933311578112} & 2436680\,$^{10L}$      & 15.187      &  --        &     --      &    --         &   --            &  --  \\
2051288445140366464 & 2569078\,$^{10L}$      & 13.650      & 4\,418(90) & 2.290(49) & -5.428(56)  & 0.200(8)      & 5  \\
2051291434437586304 & 2568912\,$^{8L}$       & 12.859      & 4\,701(95) & 3.01(6) & -7.122(55)  & -0.100(8)     & 5  \\
{\bf 2051291567575397120} & 2568656\,$^{X}$        & 15.618  &   --       &     --      &     --        &  --             &   -- \\
2051291571876502912 & 2568654\,$^{11L}$      & 13.140      & 4\,301(83) & 2.07(5)	 & -73.155(12) & 0.100(8)      & 5  \\
2051291674955780992 & 2568888\,$^{10L}$      & 14.131      & 4\,387(100)& 2.070(51) & -58.47(23)  & -0.100(11)    & 5  \\
{\bf 2051291949833615104} & 2568575\,$^{5L}$       & 13.407      & 4\,501(50) & 2.55(8)	 & -2.26(4)	   & 0.181(48)     & 3  \\
2051293118064546944 & 2437692\,$^{15L}$      & 11.203      & 4\,817(86) & 2.610(38) & -12.780(24) & 0.200(6)      & 5  \\
{\bf 2051293702180102144} & 2570002\,$^{X}$        & 13.628  &   --       &   --        &    --         &     --          &   -- \\
2051294320655654912 & 2569137\,$^{10L}$      & 13.630      & 4\,298(102)& 1.71(6) & -194.10(44) & -0.800(16)    & 5  \\
{\bf 2051295282728537216} & 2570696\,$^{2L}$       & 13.620      & 4\,461(80) & 2.13(6) & -11.38(9)   & -0.300(8)     & 5  \\
\hline
\multicolumn{8}{c}{RRLyr/BLBoo}\\
{\bf 2051294934839887872} & 2569850\,$^{X}$        & 20.426  &   --        &   --       &      --       &     --          & --\\
\hline
\multicolumn{8}{c}{semi-regular}\\
2051105032859063424 & 2438242\,$^{X}$        & 16.817      & 4\,511(99) & 2.660(47) & -47.85(8) & 0.400(9)      & 4 \\
\hline\hline
\end{tabular}
}
\end{table*}

\begin{table*}
\caption{List of rotational variables that are not cluster members. See the caption of Table\,1 for explanation.}
\label{tab:Pulsators_field}
\centering
\resizebox{\columnwidth}{320pt}{%
\begin{tabular}{clrclllll}
\hline\hline
\multicolumn{1}{c}{\multirow{2}{*}{\gaia\,EDR3}} & \multicolumn{1}{c}{\multirow{2}{*}{KIC}} & \multicolumn{1}{c}{Period} & \multicolumn{1}{c}{G} & \multicolumn{1}{c}{T$_{\rm eff}$} & \multicolumn{1}{c}{\multirow{2}{*}{\logg}} & \multicolumn{1}{c}{RV} & \multicolumn{1}{c}{\multirow{2}{*}{[Fe/H]}} & \multicolumn{1}{c}{\multirow{2}{*}{Ref}}\\
&& \multicolumn{1}{c}{[days]} & \multicolumn{1}{c}{[mag]} & \multicolumn{1}{c}{[K]} && \multicolumn{1}{c}{[km/s]} &&\\
\hline\hline
{\bf 2051098951180105088} &        --             & 1.33   & 20.929   &  --    &  --    &  --   &  -- & --\\
{\bf 2051099088624172928} & 2297416\,$^{X}$       & 2.66   & 20.201   &  --    &  --    &   --   &  -- & --\\
2051099187404218752 & 2436945\,$^{X}$        & 4.58   & 16.735   &   --   &  --    &  --    &  -- & --\\
{\bf 2051104139500783744} &      --               & 0.64   & 20.943   &   --   &  --    &   --  &  -- & --\\
2051104139505811328 & 2297846\,$^{X}$        & 13.40  & 14.901   & --     &   --   &   --   &   --& --\\
{\bf 2051104272645525760} & 2438284\,$^{X}$       & 2.63   & 19.629   &   --        &    --        &     --     &    --           & --\\
{\bf 2051104272645529472} &        --             & 1.53   & 20.277   &    --       &      --      &     --     &       --        &-- \\
2051104276944815744 & 2438272\,$^{18L}$      & 11.48  & 13.173   &   --        &   --         &     --     &   --            & --\\
2051104586182508928 & 2438513\,$^{15S,17L}$  & 13.27  & 13.939   & 5\,652(125)	& 4.51(8)	 &-35.74(33)&	0.000(10)   & 5 \\
{\bf 2051104753679401728} & 2438740\,$^{X}$       & 0.42   & 19.653   &  --         &     --       &    --      &    --           & --\\
2051104826696024832 &          --            & 1.06   & 20.370   &  --         &     --       &    --      &     --          & --\\
2051104925480543616 & 2437773\,$^{X}$        & 0.51   & 19.083   & 6\,070(110) & 2.9(1) & 8(4) & -0.54(15) & 2$^{*}$ \\
{\bf 2051105062917626624} & 2437959\,$^{X}$       & 15.27  & 17.987   & --     &   --   &    --  & --  & --\\
{\bf 2051105067218772096} & 2437888\,$^{8L}$       & 10.34  & 14.730   & 6\,120(80) & 4.93(5) & 16.42(37) & -0.39(12) & 2$^{*}$ \\
{\bf 2051105067218785280} & 2438003\,$^{18L}$      & 0.70   & 13.025   & 7\,132(30)	& 4.102(50)	 & 3(10)	& -0.22(3)    & 3 \\
{\bf 2051105135933158144} &    --                 & 3.99   & 20.627      &  --    &  --    & --     & --  & --\\
{\bf 2051105239017448448} & 2437649\,$^{X}$       & 7.62   & 16.377      &  --    &  --    &   --   &  -- & --\\
{\bf 2051105273377200384} & 2437789\,$^{X}$       & 16.61  & 17.023      &  --    &  --    &   --   &  -- & --\\
{\bf 2051105303437700224} & 2437984\,$^{X}$       & 4.70   & 19.703   &   --        &   --         &   --       &    --           &-- \\
2051105307736986496 & 2437996\,$^{17L}$      & 14.61  & 13.887   & 5\,753(37)	& 4.60(6)	 & -38(5)	& 0.104(35)     & 3 \\
2051105410811376640 & 2438305\,$^{X}$        & 7.01   & 19.692   &   --        &    --        &       --   &       --        & --\\
{\bf 2051105823133141888} & 2438376\,$^{X}$       & 11.44  & 17.788   &   --        &   --         &     --     &      --         &-- \\
{\bf 2051105926207653504} &     --                & 5.11      & 20.540   &    --       &  --          &     --     &    --           &-- \\
{\bf 2051106063646757888} & 2438861\,$^{X}$       & 1.48      & 19.672   &   --        &     --       &     --     &    --           &-- \\
2051107124503872256 & 2438516\,$^{X}$        & 5.13      & 18.313   & 5\,150(50)	& 4.773(27)	 & -36.9(15)	& 0.64(30)  & 1$^*$ \\
2051107472400670720 & 2570846\,$^{17L}$      & 10.82     & 15.313   &   --        &     --       &   --       &        --       & --\\
2051107541120138752 & 2570736\,$^{X}$        & 1.61      & 18.428   &    --       &    --        &    --      &    --           & --\\
{\bf 2051286997729896832} &     --                & 0.86      & 19.542   &   --   &   --   &   --  &   -- & --\\
2051287066448488192 &       --               & 1.62      & 19.279   & 4\,850(50)	& 4.871(40)& -112.4(1)	& -0.15(12) & 1$^{*}$ \\
2051287139470009088 & 2436635\,$^{17L}$      & 1.17   & 15.989      & 4\,445(291)	& 4.09(45) & -76(6) & -0.32(27) & 3 \\
{\bf 2051287547485746432} & 2436206\,$^{X}$       & 8.25   & 17.626      &  --    &    --  &    --  & -- & -- \\
{\bf 2051288032823472128} & 2436274\,$^{X}$       & 4.58   & 18.220      &   --   &    --  &     -- & -- & -- \\
{\bf 2051288101537860992} &     --                & 3.39   & 20.539      &  --    &   --   &    --  & --  &-- \\
{\bf 2051288165961197952} & 2436416\,$^{X}$       & 3.99   & 19.859      &  --    &   --  &  --    & -- & -- \\
{\bf 2051288445135225472} & 2569015\,$^{X}$       & 1.32   & 20.103      &  --    &   --   &    --  &--  &--  \\
{\bf 2051290919041461632} & 2435889\,$^{X}$             & 14.47  & 16.670      &  --    &  --    &  --    & -- & -- \\
{\bf 2051291434437576320} & 2568884\,$^{X}$       & 10.50  & 18.224      &  --    &  --    &    --  & -- & --\\
{\bf 2051291468792223104} &     --                  & 7.37      & 20.190   &   --   &  --    &  --   &  -- &  --\\
{\bf 2051291571876503424} & 2568672\,$^{X}$       & 18.09     & 16.477   &    --       &     --       &     --     &     --          &-- \\
2051292976335015296 & 2437180\,$^{X}$             & 4.07   & 17.907      &   --   &    --  &  --    &  -- &  --\\
2051292980625560448 & 2437292\,$^{X}$             & 5.51   & 18.356      & --     &  --    & --     &  -- & -- \\
2051292980625569792 & 2437234\,$^{17L}$           & 24.10   & 13.153     & --     &  --    &   --   &  -- & --\\
{\bf 2051293014978879744} & 2437723\,$^{X}$       & 1.17      & 18.608   &   --        &     --       &    --      &     --          & --\\
{\bf 2051293014985313152} & 2437469\,$^{X}$       & 9.85      & 19.161   &   --    &   --   &   --  &-- & --   \\
2051293014985313792 & 2437574\,$^{X}$             & 11.683    & 17.183   & 5\,650(50)	& 4.5(8)	&-45.0(3) & 0.05(7) & 1$^{*}$ \\
{\bf 2051293049338645504} &          --             & 1.97      & 20.310   &    --       &  --          &   --       &      --         & --\\
2051293083704793856 & 2437359\,$^{17L}$           & 16.91   & 14.598     &  --    &  --    &    --  & -- & --\\
{\bf 2051293152424251648} & 2436988\,$^{X}$       & 4.69   & 20.213      &  --    &   --   &   --   & -- &  --\\
{\bf 2051293221143720832} & 2436808\,$^{X}$       & 4.94   & 16.764      &  --    &  --    &  --    & -- & -- \\
{\bf 2051293255503475200} & 2436958\,$^{X}$       & 4.50   & 20.342      &  --    &  --    &   --   & -- & -- \\
2051293289863227520 & 2437329\,$^{16L}$           & 11.67     & 15.925   &   --        &    --        &    --      &      --         & --\\
2051293358577984768   &             --              & 1.55      & 20.049   &    --       &     --       &    --      &       --        & --\\
{\bf 2051293388640138880} &        --               & 8.48      & 20.028   &   --        &     --       &    --      &      --         & --\\
{\bf 2051293496015260416} & 2570182\,$^{X}$       & 1.99      & 18.719   &    --       &   --         &   --       &      --         & --\\
{\bf 2051293530375018112} &           --            & 4.27   & 20.832      &   --   &   --   &   --   &  -- & --\\
2051293599094510848 & 2570536\,$^{17L}$           & 10.03     & 14.507   &   --        &    --        &    --      &    --           & --\\
2051293839619440640 & 2570259\,$^{1S,1L}$         & 3.84      & 16.128   &   --        &    --        &    --      &    --           & --\\
{\bf 2051293908333956352} & 2570154\,$^{X}$       & 6.39      & 18.171   & 5\,380(50)	& 4.004(20)	 & -68.6(36)   & 0.12(8) & 1$^*$ \\
2051293942698672000 & 2570258\,$^{X}$           & 7.40      & 16.852   &   --        &   --         &     --     &      --         & --\\
{\bf 2051293942698685056} & 2570313\,$^{X}$       & 1.39      & 19.449   &   --        &   --         &      --    &       --        & --\\
2051293972756992384 & 2436621\,$^{X}$             & 7.58      & 18.758   &   --   &    --  & --    & -- &  -- \\
2051293977053154688 & 2436734\,$^{X}$             & 1.59      & 20.187   & --	& -- &	-- &	-- & -- \\
{\bf 2051294011418019072} & 2436852\,$^{X}$       & 3.31      & 18.871   &   --   &  --    &   --  &  -- & -- \\
{\bf 2051294080137491968} & 2569421\,$^{X}$       & 13.51  & 18.689      &   --   &    --  &  --    & -- &--  \\
\textit{\textbf{2051294114495072896}} &   --        & 3.54   & 20.960      &   --   &   --   &   --   & --  & --\\
{\bf 2051294148851973888} & 2569675\,$^{X}$       & 5.24      & 20.253   &  --         &   --         &   --       &   --            & --\\
2051294148857011712 & 2569737\,$^{14L}$           & 13.54     & 16.109   &    --       &  --          &  --        &    --           & --\\
{\bf 2051294492454406272} & 2569431\,$^{X}$       & 2.49      & 17.007   &   --        &   --         &  --        &   --            & --\\
\textit{\textbf{2051294664251895808}} &   --        & 4.90   & 21.163      &   --   & --     & --     &  -- & --\\
2051294728671468672 & 2569908\,$^{X}$           & 6.32      & 16.820   &   --        &   --         & --         &   --            & --\\
{\bf 2051294831750674944} & 2569761\,$^{X}$       & 1.99      & 20.008   &   --        &    --        &  --        &   --            & -- \\
2051294870413630208 & 2569467\,$^{X}$             & 9.78      & 15.965   &  --         &   --         & --         &  --             & --\\
2051295007845559552 &              --               & 6.08      & 20.344   &  --         &   --         & --         & --              & --\\
\textit{\textbf{2051295007845583744}} &    --       & 1.80   & 21.103      &  --    &  --    & --     & --  & --\\
{\bf 2051295488886958592} & 2570555\,$^{X}$       & 6.35      & 16.869   &  --         &    --        & --         &  --             &-- \\
{\bf 2051295557606451456} & 2570581\,$^{X}$       & 16.52     & 16.997   &  --         &  --          &   --       &  --             & --\\
2051296652817037056   & 2707961\,$^{X}$           & 8.00      & 18.949   &  --         &   --         &   --       &   --            &-- \\
\textit{2051297516106367232} &   --        & 2.65   & 21.041    &  --         &   --         &  --        & --              & --\\
{\bf 2051297962788050816} & 2707771\,$^{X}$       & 0.73      & 18.021   &   --        &   --         &   --       &   --            & --\\
{\bf 2051298031507519616} & 2707692\,$^{X}$       & 2.50      & 19.376   &  --         &   --         &  --        &    --           &-- \\
{\bf 2051299474616582912} & 2707858\,$^{X}$       & 0.91      & 20.076   &  --         &   --         &  --        &   --            & --\\
2051297275593155328 & 2569073\,$^{X}$       & 14.66  & 14.236           & --     &   --   & --     & --&  -- \\
{\bf 2051293255496998912} &        --              & 9.00      & 19.651   &  --    &   --   &  --   & --   & --\\
\hline\hline
\end{tabular}
}
\end{table*}

\begin{table*}
\caption{List of unclassified variables that are not cluster members. See the caption of Table\,1 for explanation.}
\label{tab:cluster_unclassified}
\centering
\resizebox{\columnwidth}{!}{
\begin{tabular}{cllclllll}
\hline\hline
\multicolumn{1}{c}{\multirow{2}{*}{\gaia\,EDR3}} & \multicolumn{1}{c}{\multirow{2}{*}{KIC}} & \multicolumn{1}{c}{Period} & \multicolumn{1}{c}{G} & \multicolumn{1}{c}{T$_{\rm eff}$} & \multicolumn{1}{c}{\multirow{2}{*}{\logg}} & \multicolumn{1}{c}{RV} & \multicolumn{1}{c}{\multirow{2}{*}{[Fe/H]}} & \multicolumn{1}{c}{\multirow{2}{*}{Ref}}\\
&& \multicolumn{1}{c}{[days]} & \multicolumn{1}{c}{[mag]} & \multicolumn{1}{c}{[K]} && \multicolumn{1}{c}{[km/s]} &&\\
\hline\hline
2051104654901977984 & 2438433\,$^{X}$ & 0.93 & 20.356 & -- & -- & -- & -- & --\\
{\bf 2051105307732052864} & 2438028\,$^{2L}$ & 0.06 & 20.639 & 3\,470(50) & 3.842(47) & -48.80(30) & 0.09(10) & 1$^*$ \\
{\bf 2051288337759883648} & 2436293\,$^{X}$ & 0.20 & 18.578 & -- & -- & -- & -- & --\\
\textit{\textbf{2051293736533447424}} & 2570195\,$^{X}$ & 0.77 & 20.856 & -- & -- & -- & -- & --\\
\hline
\hline\hline
\end{tabular}
}
\flushleft
\end{table*}

\begin{table*}
\caption{A list of equatorial coordinates that are associated with superstamp pixels showing signal in their amplitude spectra. In the case of the last three positions listed, we found no optical objects that could be a source of the signal.}
\label{tab:clusterunknown}
\centering
\resizebox{\columnwidth}{!}{
\begin{tabular}{cllc}
\hline\hline
\multicolumn{1}{c}{$\alpha_{\rm 2000}$} & \multicolumn{1}{c}{$\delta_{\rm 2000}$} & \multicolumn{1}{c}{Period} & \multicolumn{1}{c}{\multirow{2}{*}{Type}} \\
\multicolumn{1}{c}{[hh mm ss.ss]} & \multicolumn{1}{c}{[dd mm ss.s]} & \multicolumn{1}{c}{[days]} & \\
\hline\hline
19 20 51.72 & +37 47 45.4   & 1.10     & rotational \\
{\bf 19 20 53.56} & {\bf +37 47 05.4}   & 1.20     & eclipsing \\
{\bf 19 20 56.85} & {\bf +37 47 43.2}   & 2.40     & rotational \\
19 21 13.06 & +37 42 36.3  & 0.25     & binary \\
{\bf 19 20 31.99} & {\bf +37 47 42.4}  & 0.41     & binary \\
{\bf 19 20 45.84} & {\bf +37 47 04.1}  & 3.07     & rotational \\
{\bf 19 20 47.82} & {\bf +37 42 18.2}   & 0.68     & rotational \\
{\bf 19 21 03.71} & {\bf +37 44 17.6}   & 11.68    & rotational \\
{\bf 19 20 51.85} & {\bf +37 39 51.5}   & 9.21     & solar-like \\
\hline
19 20 32.77 & +37 43 23.8   & 0.15     & binary \\
19 20 44.47 & +37 40 51.5   & 5.73     & rotational \\
19 20 46.46 & +37 40 18.9   & 5.73     & rotational \\
\hline\hline
\end{tabular}
}
\flushleft
\end{table*}


\section{Summary}
We presented a search for variable stars in the \kep\ superstamp data. All available pixels were searched, by means of a Fourier amplitude spectrum, and a contiguous optimal aperture for each object that shows a significant flux variation were defined. The coordinates of these optimal apertures were matched with optical counterparts using Pan-STARRS. In total, we detected 278 variable stars. We cross-matched our variable star sample with those reported in the literature and found 16 variable stars reported by Kaluzny and Rucinski\,(1993), four by Rucinski \etl(1996), 23 by Mochejska \etl(2002), six by Mochejska \etl(2003), four by Kaluzny\,(2003), 15 by Bruntt \etl(2003), seven by Mochejska \etl(2005), one by Hartman \etl(2005) and 33 by de\,Marchi \etl(2007). 
We found 240 stars having KIC designations, out of which 140 stars do not have data delivered to the MAST. A variability of 119 stars was not known prior to our analysis. These stars are marked in bold in Tables\,1-8, including the first nine objects listed in Table\,9. Just recently, and independently of our work, Colman \etl(2022) reported light curves of stars with a KIC designation, however the authors did not specify which stars were found variables, only reporting a total number of 239 variables. Even accounting for this report, which is limited to the KIC stars only, we can consider 26 variable stars having no KIC designation (including seven objects marked in bold), to be new detections. No ground-based work on the detection of the variability of these 26 stars was reported, either.

Using \gaia\,EDR3 astrometry, we calculated the membership probabilities for all variable stars in our sample by applying Bayesian Gaussian mixture models. A star is considered to be a cluster member if the probability is higher than 50\%. In total, we found 129 variable stars that are cluster members, 17 binary systems, 45 pulsators, 62 rotationally and five unclassified variables. The locations of these cluster variable stars in the CMD diagram were estimated and indicate the evolutionary status of cluster members. In the CMD, a majority of our variable stars are located in the MS. Solar-like pulsators are mostly located in the RGB and RC, while semi-regular variables are located in the RGB and AGB. In addition there are five BS and three EHB stars in the CMD.

In the case of binary systems, we estimated mid-times of eclipses and derived ephemerides. We calculated the O-C diagrams and checked for any orbital period variation. Only three binary systems show significant period variation, however its nature is not confirmed. The solar-like counterpart has been a subject of a detailed analysis of its pulsation content, and the results will be published elsewhere. The analysis of three sdBVs is already reported by Sanjayan \etl(2022). The rotational variables are not subject to our detailed analysis. This type of variables can be very useful toward gyrochronology.

We utilized public and archived spectra for 111 variable stars. Spectra for 64 stars were fitted with {\sc XTgrid}, while for the remainder of the sample, we adopted the fit values from the surveys. We derived T$_{\rm eff}$, \logg, [Fe/H] and RVs. Our spectral analysis was able to recover consistent stellar parameters from very diverse spectroscopic data. This consistency is reflected by the similar distribution of stars in the CMD and HRD. The most significant limiting factor on the parameter determination was the low SNR of some spectra, while the spectral coverage and crowding in dense fields played less significant roles. We found that the metallicity is not consistent among the cluster members and it is still unclear what causes it. If the inconsistency is real a possible explanation would be the presence of multiple stellar populations within the cluster as mentioned by Geisler \etl (2012). To confirm this hypothesis a uniform spectroscopic survey is required.

MIST isochrones were fit to our CMD comprised of cluster members, including our variable star population. From our best three fits, we derived a metallicity range of 0.26 to 0.28 and the age of NGC\,6791 to be 8.91\,Gyr. Our age estimate agrees with the values reported by e.g. Choi \etl2018. The average distance estimate from the distance modulus is 4134\,pc which overlaps with our independent estimate of 4123(31)\,pc we derived from the \gaia\ astrometry.

\section*{Acknowledgement}
Financial support from the National Science Centre under projects No.\,UMO-2017/26/E/ST9/00703 and UMO-2017/25/B/ST9/02218 is acknowledged. 
PN acknowledges support from the Grant Agency of the Czech Republic 
(GA\v{C}R 22-34467S).
The Astronomical Institute in Ond\v{r}ejov is supported by the project RVO:67985815.
This paper includes data collected by the Kepler mission and obtained from the MAST data archive at the Space Telescope Science Institute (STScI). Funding for the Kepler mission is provided by the NASA Science Mission Directorate. STScI is operated by the Association of Universities for Research in Astronomy, Inc., under NASA contract NAS 5-26555. This work has made use of data from the European Space Agency (ESA) mission. 
\gaia \,(https://www.cosmos.esa.int/gaia), processed by the \gaia\,
Data Processing and Analysis Consortium (https://www.cosmos.esa.int/web/gaia/dpac/consortium). Funding for the DPAC
has been provided by national institutions, in particular the institutions
participating in the \gaia\, Multilateral Agreement.
This research has made use of the NASA/IPAC Extragalactic Database (NED),
which is operated by the Jet Propulsion Laboratory, California Institute of Technology, under contract with the National Aeronautics and Space Administration.
This research has used the services of \url{www.Astroserver.org} under reference XZR329.
We thank the anonymous referee for valuable comments, which have significantly improved the quality of the manuscript.

\section*{References}
\small
Ahn C. P. et al. 2014, ApJ, 211, 17\\
Baade W. 1931, Astronomische Nachrichten, 243, 303\\
Baran A. 2013, AcA, 63, 203\\
Basu S. et al. 2011, ApJ, 729, L10\\
Bedin L. R. et al. 2005, ApJ, 624, 45\\
Blanton M. R. et al. 2017, AJ, 154, 28\\
Bohlin, R.~C. M,\'e,sz,\'a,ros, S., Fleming, S.~W. et al. 2017, AJ, 153, 234\\
Borucki W. J. et al. 2010, Science, 327, 977\\
Bressan A. M. et al. 2012, MNRAS, 427, 127-145\\
Bryson S. T. et al. 2010, ApJ, 713, 97\\
Bruntt H., Grundahl F., Tingley B., Frandsen S., Stetson P. B., Thomsen B. 2003, A\&A, 410,323\\
Caldwell D. A. et al. 2010, ApJ, 713, 92\\
Cardelli J. A. et al. 1989, ApJ, 345, 245\\
Carraro G., Villanova S., Demarque P., McSwain M. V.and Piotto G., BedinL. R., 2006, AJ, 643, 1151\\
Carrera R. et al. 2019, A\&A, 623, A80\\
Chaboyer B., Green E. M., Liebert J. 1999, AJ, 117, 1360\\
Chambers K. C. et al. 2016, arXiv, 1612, 05560\\
Choi J., Dotter A., Conroy C., Cantiello M., Paxton B., Johnson B. D. et al. 2016,ApJ, 823, 102\\
Choi J. et al. 2018, ApJ, 863, 65\\
Colman I. L. et al. 2022, ApJS, 258, 39\\
de Marchi F. et al. 2007, A\&A, 471, 515\\
Dempster A. P., Laird N. M., Rubin D. B. 1977, Journal of the Royal Statistical Society, 39, 1\\
Dotter A. 2016, ApJS, 222, 8\\
Fabricant D. et al. 2005, PASP, 117, 1411\\
Ferguson T. S. 1973, The Annals of Statistics, 1, 209\\
Flewelling H. A. et al. 2020, ApJ, 251, 7\\
\gaia\ Collaboration et al. 2016, A\&A, 595, A1\\
\gaia\ Collaboration et al. 2021, A\&A, 649, A1\\
Geisler D. et al. 2012, ApJ, 756, L40\\
Gilmore G. et al. 2012, The Messenger, 147, 25\\
Harris W. E. and Canterna R. 1981, AJ, 86, 1332\\
Hartman J. D., Stanek K. Z., Gaudi B. S., Holman M. J., McLeod B. A. 2005, AJ, 130, 2241\\
Howell S. B. et al. 2014, PASP, 126, 398\\
Jilkova L., Carraro G., Jungwiert B., Minchev I. 2012, A\&A, 541, 64\\
Kaluzny J. and Rucinski S. M. 1993, MNRAS, 265, 34\\
Kaluzny J. and Udalski A. 1992, Acta Astronomica, 42, 29\\
Kaluzny J. 2003, Acta Astron., 53, 51\\
Kamann S., Bastian N. J., Gieles M., Balbinot E., Henault-Brunet V. 2019, MNRAS, 483, 2197-2206\\
Kinemuchi K. et al. 2012, Publications of the Astronomical Society of the Pacific, 124, 963\\
King I. R. 1964, Royal Greenwich Observatory Bulletins, 82, 106\\
Kinman T. D. 1965,ApJ,142,655\\
Koch D. G. et al. 2010, ApJ, 713, 79\\
Kwee K. and van Woerden H. 1956, Astronomical Institutes of the Netherlands, 12, 327\\
Liebert J., Saffer R. A., Green E. M. 1994, AJ, 107, 4\\
Lindegren L. et al. 2021, A\&A, 649, A4\\
Majewski S. R et al. 2017, AJ, 154, 94\\
Mochejska B. J. et al. 2005, AJ, 129, 2856\\
Mochejska B. J., Stanek K. Z., Kaluzny J. 2003, AJ, 125, 3175\\
Mochejska B. J., Stanek K. Z., Sasselov D. D., Szentgyorgyi A. H. 2002, AJ, 123, 3460\\
N,\'e,meth, P., Kawka, A., Vennes, S. 2012, MNRAS, 427, 2180\\
Pedregosa F. et al. 2011, Journal of Machine Learning Research, 12, 2825\\
Platais I. et al. 2011, AJ, 733, 1\\
Randich S. et al. 2013, The Messenger, 154, 47\\
Rucinski S. M., Kaluzny J., Hilditch R. W. 1996, MNRAS, 282, 705\\
Sanjayan S. et al. 2022, MNRAS, 509, 763-777\\
Stetson P. B., Bruntt H., Grundahl F. 2003, PASP, 115, 413\\
Thompson, S, Fraquelli, D., Van Cleve, J., and Caldwell, D. 2016,KDMC, 10008,006\\
Tofflemire B. M., Gosnell N. M., Mathieu R. D., Platais I. 2014, AJ, 148, 61\\
Villanova S., Carraro G., Geisler D., Monaco L., Assmann P. 2018, ApJ, 867, 34\\
Zhao G.et al. 2012, A\&A, 12, 7\\Ahn C. P. et al. 2014, ApJ, 211, 17\\
Baade W. 1931, Astronomische Nachrichten, 243, 303\\
Baran A. 2013, AcA, 63, 203\\
Basu S. et al. 2011, ApJ, 729, L10\\
Bedin L. R. et al., 2005, ApJ, 624, 45\\
Blanton M. R. et al., 2017, AJ, 154, 28\\

\end{document}